\documentclass[useAMS,usenatbib]{mn2e}
\usepackage{amsmath,amssymb,graphicx}

\newcommand{\Msun}{\ensuremath{\,{M}_\odot}}                      
\newcommand{\Rsun}{\ensuremath{\,{R}_\odot}}                      
\newcommand{\Teff}{\ensuremath{T_{\rm eff}}}                      
\newcommand{\Mjup}{\ensuremath{\,{M}_{\rm Jup}}}                  
\newcommand{\Rjup}{\ensuremath{\,{R}_{\rm Jup}}}                  
\newcommand{\Teq}{\ensuremath{T_{\rm eq}^{\,\prime}}}             
\newcommand{\safronov}{\ensuremath{\Theta}}                       
\newcommand{\kms}{\,km\,s$^{-1}$}                                 
\newcommand{\ms}{\,m\,s$^{-1}$}                                   
\newcommand{\mss}{\,m\,s$^{-2}$}                                  
\newcommand{\as}{\ensuremath{^{\prime\prime}}}                    
\newcommand{\am}{\ensuremath{^\prime}}                            
\newcommand{\FeH}{\ensuremath{\left[\frac{\rm Fe}{\rm H}\right]}} 
\newcommand{\pjup}{\ensuremath{\,\rho_{\rm Jup}}}                 
\newcommand{\psun}{\ensuremath{\,\rho_\odot}}                     
\newcommand{\mc}[1]{\multicolumn{2}{c}{#1}}
\newcommand{\mcc}[1]{\multicolumn{3}{c}{#1}}

\newcommand{\erc}[3]{\mc{\ensuremath{#1^{+#2}_{-#3}}}}

\newcommand{\ermcc}[5]{\mcc{\ensuremath{{#1\,^{+#2}_{-#3}}\,^{+#4}_{-#5}}}}

\title[Physical properties of WASP-19b]
{Physical properties, transmission and emission spectra of the
WASP-19 planetary system from multi-colour photometry\thanks{Based
on data collected with the Gamma Ray Burst Optical and
Near-Infrared Detector (GROND) at the MPG/ESO 2.2\,m telescope and
by MiNDSTEp with the Danish 1.54\,m telescope at the ESO
Observatory in La Silla.}}

\author[L.~Mancini et al.]
{\parbox{\textwidth}{L.~Mancini$^{1,2}$\thanks{E-mail:
\texttt{mancini@mpia.de}}, %
        S.\ Ciceri\,$^{1}$,
        G.\ Chen\,$^{1,3}$,
        J.\ Tregloan-Reed\,$^{4}$,
    J.\,J.\ Fortney\,$^{5}$,
        J.\ Southworth\,$^{4}$,
        T.G.\ Tan\,$^{6}$,
        M.\ Burgdorf\,$^{7}$,
        S.\ Calchi Novati\,$^{8,2}$,
        M.\ Dominik\,$^{9}$\thanks{Royal Society University Research Fellow},
        X.-S.\ Fang\,$^{10}$,
        F.\ Finet\,$^{11}$,
        T.\ Gerner\,$^{12}$,
        S.\ Hardis\,$^{13,14}$,
        T.\ C.\ Hinse\,$^{15}$,
        U.\ G.\ J{\o}rgensen\,$^{13,14}$,
        C.\ Liebig\,$^{9}$,
        N.\ Nikolov\,$^{1,16}$,
        D.\ Ricci\,$^{11,17}$,
        S.\ Sch\"afer\,$^{18}$,
        F.\ Sch\"onebeck\,$^{12}$,
        J.\ Skottfelt\,$^{13,14}$,
        O.\ Wertz\,$^{11}$,
    K.\,A.\ Alsubai\,$^{19}$,
        V.\ Bozza\,$^{2,20}$,
        P.\ Browne\,$^{10}$,
        P.\ Dodds\,$^{9}$,
     S.-H.\ Gu\,$^{10,21}$,
        K.\ Harps{\o}e\,$^{13,14}$,
        Th.\ Henning\,$^{1}$,
        M.\ Hundertmark\,$^{9}$,
        J.\ Jessen-Hansen\,$^{22}$,
        N.\ Kains\,$^{23}$,
        E.\ Kerins\,$^{24}$,
        H.\ Kjeldsen\,$^{22}$,
    M.\,N.\ Lund\,$^{22}$,
        M.\ Lundkvist\,$^{22}$,
        N.\ Madhusudhan\,$^{25}$,
        M.\ Mathiasen\,$^{13,14}$,
    M.\,T.\ Penny\,$^{26}$,
        S.\ Proft\,$^{12}$,
        S.\ Rahvar\,$^{27}$,
        K.\ Sahu\,$^{28}$,
        G.\ Scarpetta\,$^{8,2,20}$,
        C.\ Snodgrass\,$^{29}$ and
        J.\ Surdej\,$^{11}$} \vspace{0.4cm}\\
\parbox{\textwidth}{
$^{1}$\,Max Planck Institute for Astronomy, K\"onigstuhl 17, 69117 Heidelberg, Germany \\
        $^{2}$\,Department of Physics, University of Salerno, 84084-Fisciano (SA), Italy \\
        $^{3}$\,Purple Mountain Observatory \& Key Laboratory for Radio Astronomy, 2 West Beijing Road, Nanjing 210008, China \\
        $^{4}$\,Astrophysics Group, Keele University, Staffordshire, ST5 5BG, UK \\
        $^{5}$\,Department of Astronomy \& Astrophysics, University of California, Santa Cruz, CA 95064, USA \\
        $^{6}$\,Perth Exoplanet Survey Telescope, Perth, Australia \\
        $^{7}$\,HE Space Operations GmbH, Flughafenallee 24, D-28199 Bremen, Germany \\
        $^{8}$\,Istituto Internazionale per gli Alti Studi Scientifici (IIASS), 84019 Vietri Sul Mare (SA), Italy \\
        $^{9}$\,SUPA, University of St Andrews, School of Physics \& Astronomy, North Haugh, St Andrews, KY16 9SS, UK \\
        $^{10}$\,National Astronomical Observatories/Yunnan Observatory, Chinese Academy of Sciences, Kunming 650011, China \\
        $^{11}$\,Institut d'Astrophysique et de G\'eophysique, Universit\'e de Li\`ege, 4000 Li\`ege, Belgium \\
        $^{12}$\,Zentrum f\"ur Astronomie, Universit\"at Heidelberg, M\"onchhofstra{\ss}e 12-14, 69120 Heidelberg, Germany \\
        $^{13}$\,Niels Bohr Institute, University of Copenhagen, Juliane Maries vej 30, 2100 Copenhagen \O, Denmark \\
        $^{14}$\,Centre for Star and Planet Formation, Geological Museum, {\O}ster Voldgade 5-7, 1350 Copenhagen, Denmark \\
        $^{15}$\,Korea Astronomy and Space Science Institute, Daejeon 305-348, Republic of Korea \\
        $^{16}$\,Astrophysics Group, University of Exeter, Stocker Road, EX4 4QL, Exeter, UK \\
        $^{17}$\,Instituto de Astronom\'{i}a - UNAM, Km 103 Carretera Tijuana Ensenada, 422860, Ensenada (Baja Cfa), Mexico \\
        $^{18}$\,Institut f\"ur Astrophysik, Georg-August-Universit\"at G\"ottingen, Friedrich-Hund-Platz 1, 37077 G\"ottingen, Germany \\
        $^{19}$\,Qatar Foundation, PO Box 5825, Doha, Qatar \\
        $^{20}$\,Istituto Nazionale di Fisica Nucleare, Sezione di Napoli, Napoli, Italy \\
        $^{21}$\,Key Laboratory for the Structure and Evolution of Celestial Objects, Chinese Academy of Sciences, Kunming 650011, China \\
        $^{22}$\,Stellar Astrophysics Centre, Dep.\ of Physics and Astronomy, Aarhus University, Ny Munkegade 120, 8000 Aarhus C, Denmark \\
        $^{23}$\,European Southern Observatory, Karl-Schwarzschild-Stra{\ss}e 2, 85748 Garching bei M\"unchen, Germany \\
        $^{24}$\,Jodrell Bank Centre for Astrophysics, University of Manchester, Oxford Road, Manchester M13 9PL, UK \\
        $^{25}$\,Department of Physics and Department of Astronomy, Yale University, New Haven, CT 06511, USA \\
        $^{26}$\,Department of Astronomy, Ohio State University, 140 W. 18th Ave., Columbus, OH 43210, USA \\
        $^{27}$\,Department of Physics, Sharif University of Technology, P.\,O.\,Box 11155-9161 Tehran, Iran \\
        $^{28}$\,Space Telescope Science Institute, 3700 San Martin Drive, Baltimore, MD 21218, USA \\
        $^{29}$\,Max-Planck-Institute for Solar System Research, Max-Planck Str.\ 2, 37191 Katlenburg-Lindau, Germany\\
}}

\begin{document} %
\maketitle
\clearpage

\begin{abstract}
We present new ground-based, multi-colour, broad-band photometric
measurements of the physical parameters, transmission and emission
spectra of the transiting extrasolar planet WASP-19\,b. The
measurements are based on observations of eight transits and four
occultations through a Gunn $i$ filter using the 1.54\,m Danish
Telescope, 14 transits through an $R_c$ filter at the PEST
observatory and one transit observed simultaneously through four
optical (Sloan $g^{\prime}$, $r^{\prime}$, $i^{\prime}$,
$z^{\prime}$) and three near-infrared ($J$, $H$, $K$) filters,
using the GROND instrument on the MPG/ESO 2.2m telescope. The
GROND optical light curves have a point-to-point scatter around
the best-fitting model between 0.52 and 0.65 mmag rms. We use
these new data to measure refined physical parameters for the
system. We find the planet to be more bloated
($R_{\mathrm{b}}=1.410 \pm 0.017$\Rjup; $M_{\mathrm{b}}=1.139 \pm
0.030$\Mjup) and the system to be twice as old as initially
thought. We also used published and archived data sets to study
the transit timings, which do not depart from a linear ephemeris.
We detected an anomaly in the GROND transit light curve which is
compatible with a spot on the photosphere of the parent star. The
starspot position, size, spot contrast and temperature were
established. Using our new and published measurements, we
assembled the planet's transmission spectrum over the
$370$--$2350$\,nm wavelength range and its emission spectrum over
the $750$--$8000$\,nm range. By comparing these data to
theoretical models we investigated the theoretically predicted
variation of the apparent radius of WASP-19\,b as a function of
wavelength and studied the composition and thermal structure of
its atmosphere. We conclude that: ($i$) there is no evidence for
strong optical absorbers at low pressure, supporting the common
idea that the planet's atmosphere lacks a dayside inversion;
($ii$) the temperature of the planet is not homogenized, because
the high warming of its dayside causes the planet to be more
efficient in re-radiating than redistributing energy to the night
side; ($iii$) the planet seems to be outside of any current
classification scheme.
\end{abstract}


\section{Introduction}
\label{sec:1}
The study of extrasolar planets is now decisively in a phase of
rapid expansion and extraordinary discoveries. More than 850
planets have been detected orbiting around stars other than our
Sun. They show a wide range of characteristics in terms of mass,
radius, density, temperature, and distance from their parent
stars. Among these planets, those that transit are prime sources
of information on the physical properties and atmospheric
characteristics of planets. Such information is precious as we
seek to establish reliable theories of their formation and
evolution. For these reasons, transiting extrasolar planets (TEPs)
are the most important and interesting ones to study in detail.

\begin{figure}%
\includegraphics[width=8.5cm]{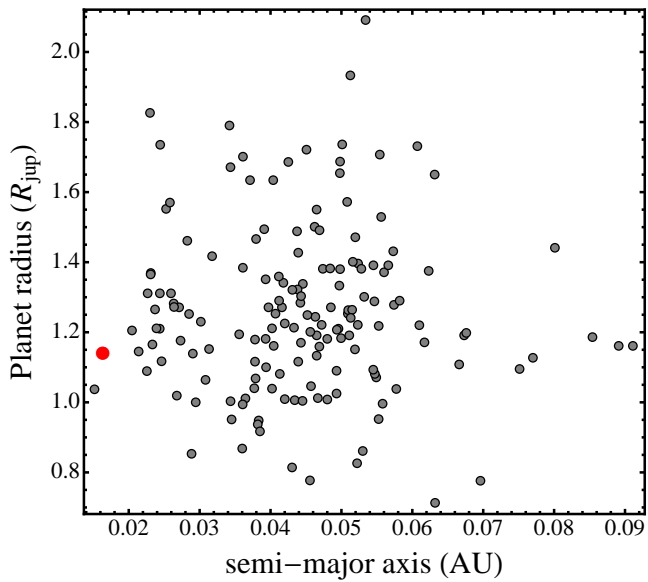}
\caption{\label{fig:00} Plot of planet radius versus semimajor
axis for the known TEPs with mass $>0.1\, M_{\mathrm{jup}}$ and
$P_{\mathrm{orb}}<10$ days. The red point represents WASP-19b. The
error bars of the planets have been suppressed for clarity. Data
taken from Transiting Extrasolar Planet Catalogue, available at
http://www.astro.keele.ac.uk/jkt/tepcat/.}
\end{figure}

In this paper we focus on WASP-19\,b \citep{hebb2010}, which is an
extremely short period TEP moving on a circular orbit
\citep{anderson2011,lendl2012} around a G8\,V star every 0.79
days. \citet{hebb2010} measured its mass to be $1.15 \pm 0.08 \,
M_{\mathrm{Jup}}$ and its radius to be $1.31 \pm 0.06 \,
R_{\mathrm{Jup}}$, which together imply that WASP-19\,b is a
slightly bloated hot Jupiter. Its orbital semimajor axis, $\sim
0.016$\,AU, places it in the class of highly irradiated planets.
The star-planet separation is equivalent to only 1.2 times the
Roche tidal radius and suggests that WASP-19\,b has spiralled
inward to its current location likely via the Kozai mechanism or
tidal dissipation \citep{hellier2011}. Its position in the
planetary radius-separation diagram is shown in
Fig.\,\ref{fig:00}.

Radial velocity (RV) follow-up observations of WASP-19 have been
obtained using the HARPS spectrograph to successfully detect the
Rossiter-McLaughlin effect. This resulted in a measurement of the
angle between the planet's orbit and the sky-projected stellar
rotation axis of $\lambda=4.6^{\circ} \pm 5.2^{\circ}$, implying
an aligned orbit for WASP-19\,b \citep{hellier2011}. Both
photometric and spectroscopic data show modulations compatible
with starspot activity on the parent star
\citep{hebb2010,hellier2011,abe2013}. This activity has been
directly observed by \citet{tregloan2013}, who reported a starspot
in two consecutive planetary transits of WASP-19\,b. They used it
to obtain a better constraint on $\lambda$ ($1.0^{\circ} \pm
1.2^{\circ}$) and measure the star's rotation period and velocity
($11.76 \pm 0.09$ d and $3.88 \pm 0.15$ km\,s$^{-1}$,
respectively).

Occultations (secondary eclipses) of WASP-19\,b, observed in
several near-infrared (NIR) \citep{gibson2010,anderson2010} and
infrared (IR) \citep{anderson2011} bands with the VLT and Spitzer
respectively, suggest that the dayside atmosphere of WASP-19\,b
lacks a temperature inversion. This picture does not support the
``TiO/VO hypothesis'' of \citet{fortney2008}, and is instead in
favour of a monotonically decreasing atmospheric temperature with
pressure. Evidence of WASP-19\,b occultations from the ground at
optical wavelengths have also been presented by
\citet{burton2012,lendl2012} and \citet{abe2013}, extending our
knowledge of the thermal emission of WASP-19\,b. Recently, time
series spectroscopy in nine channels (1250--2350\,nm) were
performed by \citet{bean2013} on the Magellan II telescope during
two transits and two occultations of WASP-19. These observations
again question a thermal inversion and favour the presence of HCN
and H$_{2}$O molecules. The whole set of data occultation slighlty
supports a carbon-rich instead of an oxygen-rich planetary
atmosphere \citep{madhusudhan2012}.

In this paper we present observations of eight transits and four
occultations in the WASP-19 system obtained with the 1.54\,m
Danish Telescope, one transit observed simultaneously in four
optical and three NIR passbands using GROND on the MPG/ESO 2.2\,m
telescope, and 14 transits observed with the 0.3\,m Perth
Exoplanet Survey Telescope (PEST). These new data allow us to
study the physical properties of the whole system, and the
transmission and emission spectra of the planet. The differential
photometry will be made available at the CDS.

Our paper is structured as follows. In Sect.\,\ref{sec:2} we
describe the observations and data reduction methodology. In
Sect.\,\ref{sec:3} we analyse the transit light curves to refine
the orbital ephemerides, obtain the photometric parameters of the
WASP-19 system, and characterize the starspot detected in the
GROND light curves. The revision of the physical properties of the
system is presented in Sect.\,\ref{sec:4}, while
Sect.\,\ref{sec:5} is dedicated to the study of the variation of
the planetary radius as function of wavelength. The analysis of
the occultation light curves is described in Sect.\,\ref{sec:6}.
Our conclusions are summarized in Sect.\,\ref{sec:7}.

\section{Observations and data reduction}
\label{sec:2}

\subsection{Transits}
\label{sec:2.1}

\begin{figure*}%
\includegraphics[width=18.0cm]{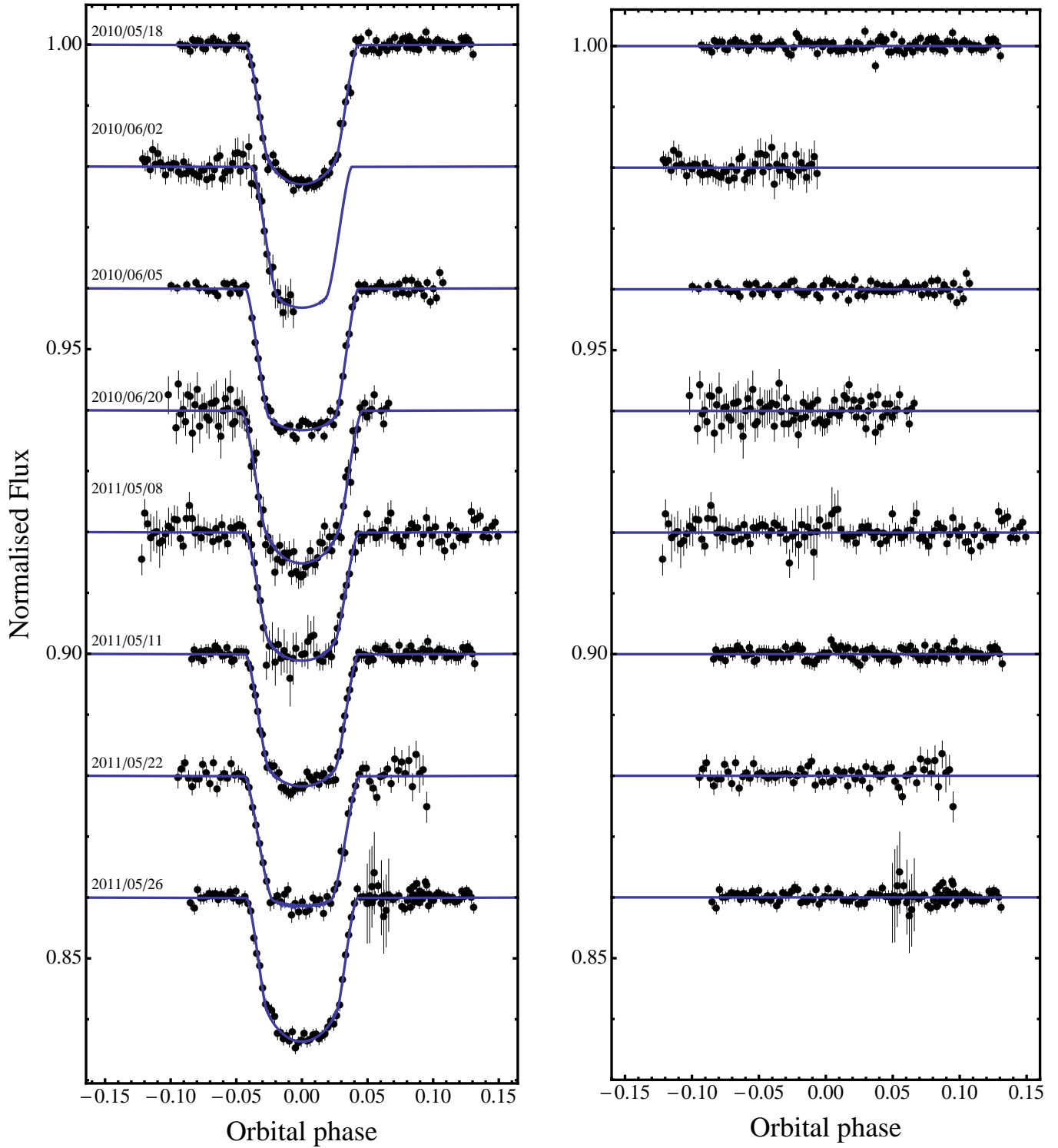}
\caption{\label{fig:01} Light curves of eight transits observed
with DFOSC through a Gunn $i$ filter (\emph{left panel}), plotted
in date order. The {\sc jktebop} best fit is also shown for each
data set, and the residuals of each fit are plotted in the
\emph{right panel}.}
\end{figure*}
%
\begin{figure*} %
\includegraphics[width=18.0cm]{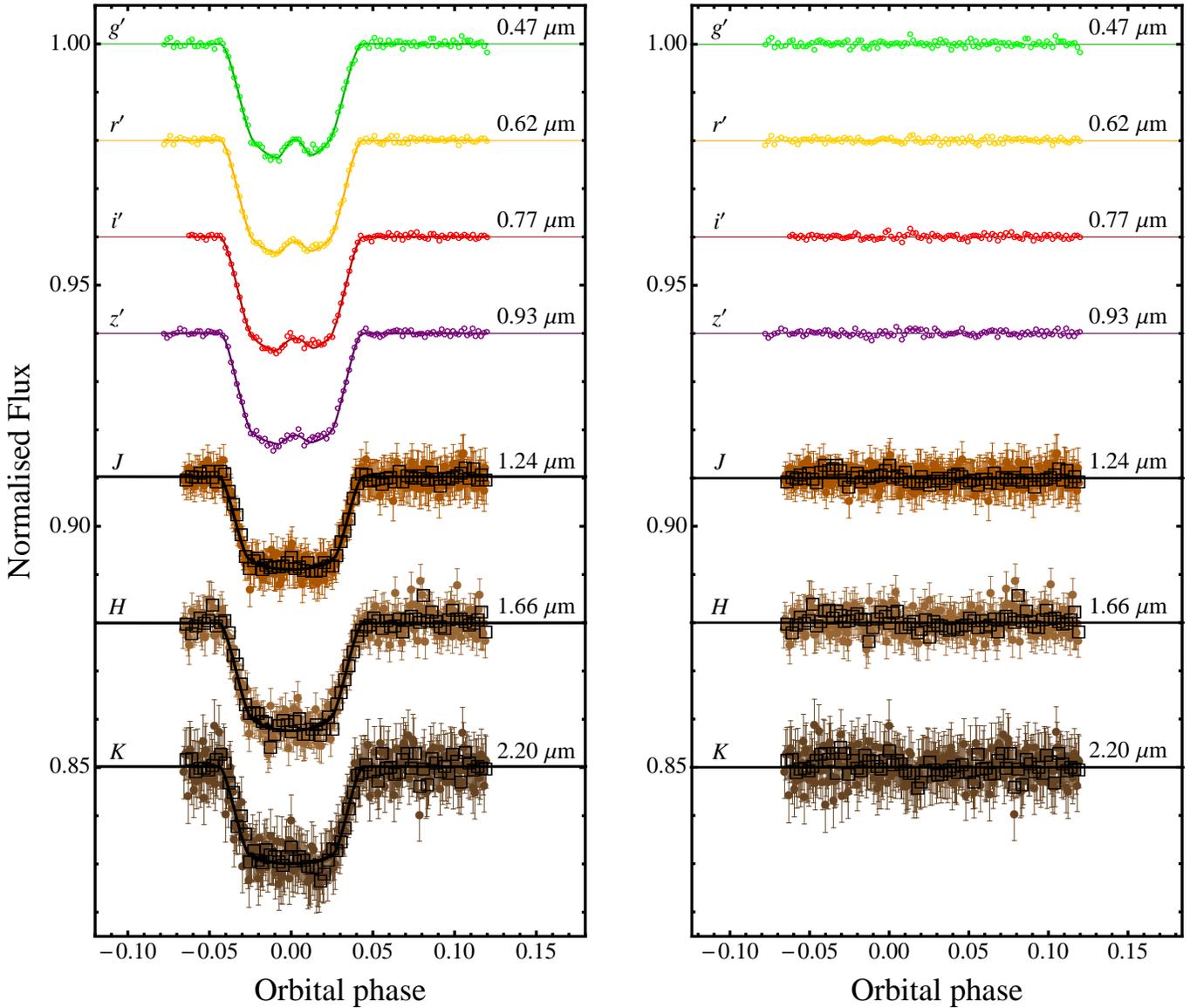}
\caption{\label{fig:02} \emph{Left-hand panel:} simultaneous
optical and NIR light curves of one transit event of WASP-19\,b
observed with GROND. Black empty squares show the three NIR light
curves binned using a bin size of 0.002 phase units. The {\sc
prism+gemc} best fits are shown as solid lines for each optical
data set, while for the NIR ones the best fits were found with
{\sc jktebop}. The passbands are labelled on the left of the
figure, and their central wavelengths are given on the right. The
error bars of the optical data have been suppressed for clarity.
The bump observed in the optical light curves, roughly at the
mid-transit time, is interpreted as the occultation of a starspot
by the planet (see Sect.\,\ref{sec:3.3}). \emph{Right-hand panel:}
the residuals of each fit.}
\end{figure*}
%

\begin{table*} \centering
\caption{\label{tab:01} Details of the transit observations
presented in this work. $N_{\rm obs}$ is the number of
observations, $T_{\rm exp}$ is the exposure time, $T_{\rm obs}$ is
the observational cadence, and  `Moon illum.' is the fractional
illumination of the Moon at the midpoint of the transit.}
\setlength{\tabcolsep}{4pt}
\begin{tabular}{lccccccccccc} \hline
Transit & Date of   & Start time & End time  &$N_{\rm obs}$ & $T_{\rm exp}$ & $T_{\rm obs}$ & Filter & Airmass & Moon & Aperture   & Scatter \\

        & first obs &    (UT)    &   (UT)    &              & (s)           & (s)           &        &         &illum.& radii (px) & (mmag)  \\
\hline
DFOSC & 2010 05 18 & 01:04 & 04:37 &  95 & 100 & 125 & Gunn $i$ & 1.15 $\to$ 2.49 & 0.19 & 17, 40, 65  & 0.83 \\
DFOSC & 2010 06 02 & 00:20 & 03:22 &  75 & 90  & 115 & Gunn $i$ & 1.19 $\to$ 2.27 & 0.76 & 18, 35, 65  & 1.33 \\
DFOSC & 2010 06 05 & 23:28 & 03:24 &  80 & 120 & 155 & Gunn $i$ & 1.12 $\to$ 2.53 & 0.39 & 19, 40, 65  & 0.88 \\
DFOSC & 2010 06 20 & 23:08 & 02:18 &  92 & 90  & 115 & Gunn $i$ & 1.19 $\to$ 2.45 & 0.70 & 16, 40, 65  & 1.92 \\
DFOSC & 2011 05 08 & 00:04 & 05:12 & 121 & 120 & 155 & Gunn $i$ & 1.05 $\to$ 2.43 & 0.22 & 16, 35, 45  & 1.60 \\
DFOSC & 2011 05 11 & 23:16 & 03:31 & 121 & 90  & 115 & Gunn $i$ & 1.05 $\to$ 1.65 & 0.64 & 17, 35, 65  & 0.80 \\
DFOSC & 2011 05 22 & 23:47 & 03:53 & 71  & 150 & 185 & Gunn $i$ & 1.09 $\to$ 2.19 & 0.67 & 24, 45, 75  & 1.43 \\
DFOSC & 2011 05 26 & 23:03 & 02:38 & 82  & 120 & 155 & Gunn $i$ & 1.05 $\to$ 1.62 & 0.29 & 19, 35, 65  & 0.99 \\
GROND & 2012 04 15 & 23:17 & 03:02 & 114 & 90  & 120 & Gunn $g^{\prime}$ & 1.08 $\to$ 1.00 $\to$ 1.15 & 0.23 & 33, 55, 85  & 0.59 \\
GROND & 2012 04 15 & 23:17 & 03:02 & 114 & 90  & 120 & Gunn $r^{\prime}$ & 1.08 $\to$ 1.00 $\to$ 1.15 & 0.23 & 33, 55, 80  & 0.52 \\
GROND & 2012 04 15 & 23:17 & 03:02 & 114 & 90  & 120 & Gunn $i^{\prime}$ & 1.08 $\to$ 1.00 $\to$ 1.15 & 0.23 & 36, 55, 80 & 0.52 \\
GROND & 2012 04 15 & 23:17 & 03:02 & 114 & 90  & 120 & Gunn $z^{\prime}$ & 1.08 $\to$ 1.00 $\to$ 1.15 & 0.23 & 32, 55, 80 & 0.65 \\
GROND & 2012 04 15 & 23:17 & 03:02 & 339 & 10  & 50 & $J$ & 1.08 $\to$ 1.00 $\to$ 1.15 & 0.23 & 6.5, 10.5, 21 & 1.59 \\
GROND & 2012 04 15 & 23:17 & 03:02 & 339 & 10  & 50 & $H$ & 1.08 $\to$ 1.00 $\to$ 1.15 & 0.23 & 5, 12, 22 & 2.22 \\
GROND & 2012 04 15 & 23:17 & 03:02 & 339 & 10  & 50 & $K$ & 1.08 $\to$ 1.00 $\to$ 1.15 & 0.23 & 7, 11, 19 & 2.82 \\
\hline \end{tabular} \end{table*}

One partial transit and seven full transits of WASP-19 were
observed using the DFOSC imager mounted on the 1.54\,m Danish
Telescope at La Silla during the MiNDSTEp campaigns in 2010 and
2011 \citep{dominik2010}. The instrument has a field of view (FOV)
of 13.7\am$\times$13.7\am\ and a plate scale of
0.39\as\,pixel$^{-1}$. All the observations were performed through
a Gunn $i$ filter and using the \emph{defocussing} method
\citep{southworth2009a,southworth2009b}. The telescope was
autoguided and the CCD was windowed to reduce the readout time.
Night logs are reported in Table\,\ref{tab:01}. The data were
analysed using the {\sc idl}\footnote{The acronym {\sc idl} stands
for Interactive Data Language and is a trademark of ITT Visual
Information Solutions. For further details see {\tt
http://www.ittvis.com/ProductServices/IDL.aspx}.} pipeline from
\citet{southworth2009a}, which uses the {\sc idl} implementation
of {\sc daophot} \citep{stetson1987}. The images were debiased and
flat-fielded using standard methods, then subjected to aperture
photometry using the {\sc aper}\footnote{{\sc aper} is part of the
{\sc astrolib} subroutine library distributed by NASA. For further
details see {\tt http://idlastro.gsfc.nasa.gov}.} task and an
optimal ensemble of comparison stars. Pointing variations were
followed by cross-correlating each image against a reference
image. The shape of the light curves is very insensitive to the
aperture sizes, so we chose those which yielded the lowest
scatter. The absolute values of the experimental errors calculated
by {\sc aper} are often underestimated, so we enlarged the error
bars for each transit to give a reduced $\chi^{2}$ of
$\chi_{\nu}^{2}=1$. We then further inflated the error bars using
the $\beta$ approach (e.g.\
\citealt{gillon2006,winn2008,gibson2008}) to account for any
correlated noise. The eight light curves are plotted in
Fig.\,\ref{fig:01}.

One transit of WASP-19 was observed on 2012 April 15, with the
\textbf{G}amma \textbf{R}ay Burst \textbf{O}ptical and
\textbf{N}ear-Infrared \textbf{D}etector (GROND) instrument
mounted on the MPG\footnote{Max Planck Gesellschaft.}/ESO 2.2\,m
telescope at ESO La Silla, Chile. GROND is an imaging system
capable of simultaneous photometric observations in four optical
(similar to Sloan $g^{\prime}$, $r^{\prime}$, $i^{\prime}$,
$z^{\prime}$) and three NIR ($J,\, H,\, K$) passbands
\citep{greiner2008}. Each of the four optical channels is equipped
with a back-illuminated $2048 \times 2048$ E2V CCD, with an FOV of
$5.4^{\prime} \times 5.4^{\prime}$ at a scale of
$0.158^{\prime\prime}/\rm{pixel}$. The three NIR channels use
$1024 \times 1024$ Rockwell HAWAII-1 arrays with a FOV of
$10^{\prime}\times 10^{\prime}$ at
$0.6^{\prime\prime}/\rm{pixel}$. The accuracy of the GROND
photometry was recently analysed by \citet{pierini2012}, who
computed the uncertainty of the total photometry in each of the
broad-bands as a function of the total magnitude of sources with
different brightness. From this analysis it emerges that for
bright stars ($<15$ mag in each passband) the uncertainty in the
optical bands is $<0.01$\,mag, while that in the NIR bands is
$\geqslant 0.02$\,mag, the $K$ band giving the worst results. In
order to improve the accuracy, we used the telescope defocussing
technique during the follow-up of the planetary transit of
WASP-19, which indeed provided a better photometry (by a factor
$\sim$3) in comparison to previous uses of this instrument (see
\citealt{nikolov2012,harpsoe2013,mancini2013b}).

The optical data were reduced as for the Danish Telescope. The NIR
data were calibrated by performing dark subtraction, readout
pattern removal and flat division on the raw science images. Both
the dark and flat master frames were created through
median-combining a stack of individual measurements, for which the
twilight sky flats were adopted. We also removed the odd-even
readout pattern (along the $x$-axis) before the flat correction by
shifting the count level of each column to a common overall level.
After the calibration, we obtained light curves using an aperture
photometry routine again based on the {\sc idl/daophot} package.
To determine the centroids of WASP-19 and its comparison stars, we
smoothed the defocused images into a lower resolution and fitted
Gaussians to the marginal $x$, $y$ distributions. We also recorded
the full-width at half-maximum (FWHM) by fitting the wing of
original PSF, the central region of which has been masked out. We
carefully chose the comparison stars that were of similar
brightness and well within the saturation limit. The companions
close to our chosen stars were masked out in case they
contaminated the photometry. Fluxes of these comparison stars were
weighted-averaged together after self-normalization, and divided
from that of WASP-19. We also experimented with a set of aperture
and sky annulus sizes, and adopted that which yielded the light
curve with least scatter. The final light curves were decorrelated
with star position, FWHM, time and airmass. Uncertainties were
treated as for the optical case. Optical and NIR GROND light
curves are shown in Fig.\,\ref{fig:02}. A clear anomaly is visible
in the four optical bands close to the midpoint of the transit.

\begin{figure} %
\includegraphics[width=0.48\textwidth,angle=0]{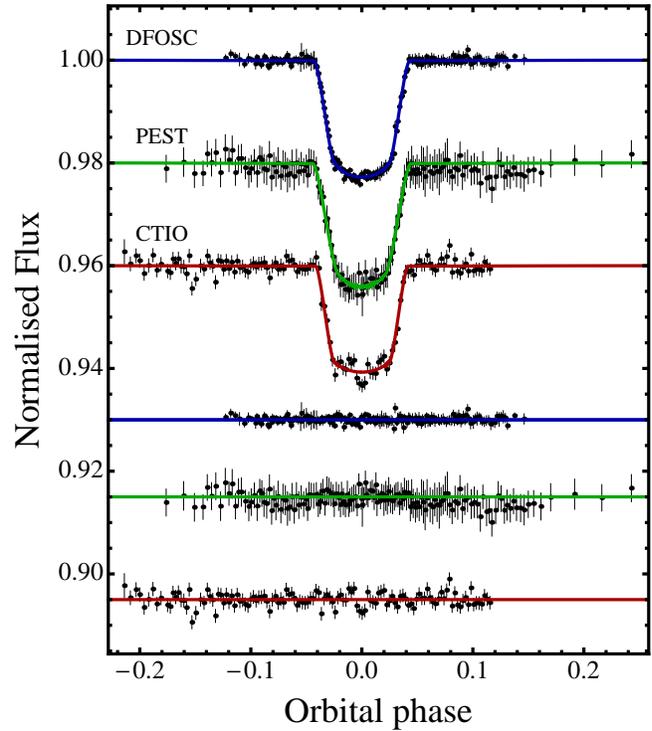}
\caption{\label{fig:03} \emph{Top}: transit light curve of WASP-19
obtained by phase-binning (bin size 0.01) all the DFOSC light
curves shown in Fig.\,\ref{fig:01}. \emph{Middle}: phased-binned
light curve derived from the 14 transits observed with PEST.
\emph{Bottom}: light curve one transit observed by
\citet{dragomir2011} using the CTIO 1\,m telescope and a Cousins
$R$ filter. The {\sc jktebop} best fit is also shown for each data
set, and the residuals of each fit are plotted near the base of
the figure.}
\end{figure}

Fourteen transits of WASP-19 were observed between 2011 February
and November at the PEST observatory, which consists of a 30\,cm
Meade LX200 SCT f/10 telescope, with a focal reducer converting
this to f/5, and equipped with an SBIG ST-8XME camera. The image
scale is 1.2 arcsec pixel$^{-1}$ and the FOV is $31 \times 21$
arcmin$^{2}$. This backyard observatory is located in a suburb of
the city of Perth (Western Australia). The transit observations
were performed using a Astrodon $R_c$ filter and the telescope was
autoguided. The images were reduced and the photometry performed
using the software package C-Munipack. Raw sky-flat frames were
bias-subtracted and then median combined to obtain a master flat
frame. For each data set, sky flats from the dusk preceding, or
dawn succeeding the observations were used when possible,
otherwise the latest available master flat was used. The science
frames were dark subtracted and flat-field corrected and then
subjected to aperture photometry. Finally, all the data sets were
phased-binned and the corresponding light curve is plotted in
Fig.\,\ref{fig:03}.

%

\subsection{Occultations}
\label{sec:2.2}
\begin{figure*} %
\includegraphics[width=18.0cm]{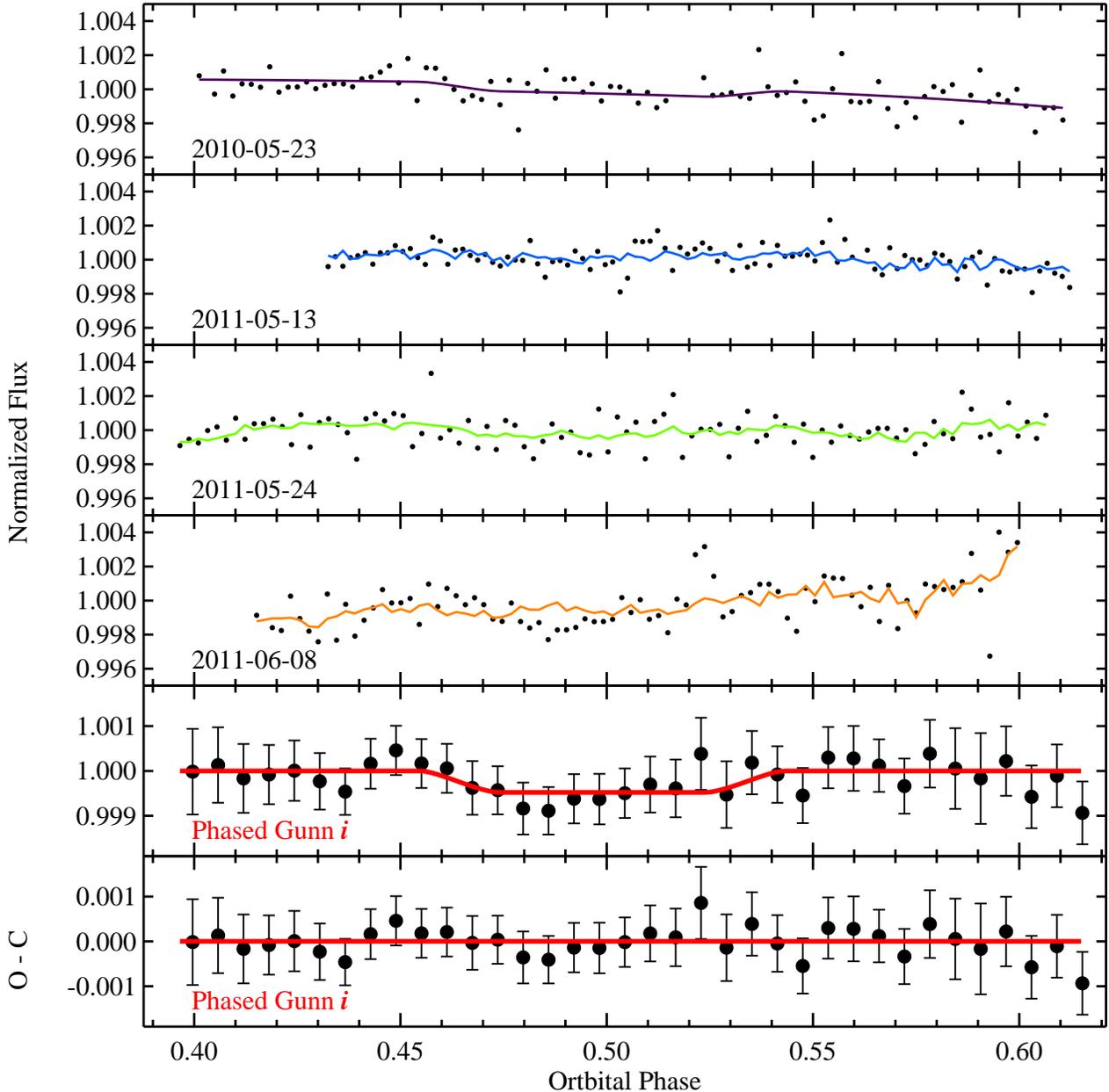}
\caption{\label{fig:04} Occultation light curves observed with
DFOSC in Gunn $i$. {\it Panels 1 to 4:} the raw light curves of
four individual nights, overplotted with the best-fitting model.
{\it Panel 5:} the phase-folded occultation light curve after
correction for baseline slope and binned in every 7 minute. {\it
Panel 6:} corresponding O--C residuals.}
\end{figure*}

Four occultations of WASP-19 were followed through a Gunn $i$
filter using the Danish Telescope and DFOSC. One was observed in
2010 May and the others between 2011 May and June
(Table\,\ref{tab:02}). The images were calibrated in the standard
way, including bias subtraction and flat division. We created
master bias and master skyflat frames through median combination
of a set of individual files, and then corrected the bias and
flat-field from the science images. We performed aperture
photometry on the calibrated science images employing the {\sc
idl/daophot} routines. Since the telescope was heavily defocused,
to find the centroids of stars, we convolved the science images
with a Gaussian kernel so that the doughnut-shaped PSFs became
approximately Gaussian. The locations of WASP-19 as well as an
ensemble of nearby comparison stars of similar brightness were
determined by employing {\sc idl/find} on these convolved images.
We placed 30 apertures on each star in steps of 1 pixel. 10 annuli
for each aperture were tried in steps of 2 pixels. We normalized
the light curve of each star by dividing their out-of-eclipse flux
level individually. We checked each star to remove the ones that
were saturated and carefully chose the ensemble with a similar
light curve shape to WASP-19. The chosen ensemble was
weight-combined as the composite reference light curve, which was
then used to normalize the WASP-19 light curve. These normalized
light curves were modelled, and the one with the lowest standard
deviation of the residuals was chosen as the optimal light curve.
The aperture settings are listed in Table\,\ref{tab:02} together
with the details of the observations. The four light curves are
shown in Fig.\,\ref{fig:04}

\begin{table*} \centering
\caption{\label{tab:02} Details of the occultation observations
presented in this work. See Table\,\ref{tab:02} for explanation of
$N_{\rm obs}$, $T_{\rm exp}$, $T_{\rm obs}$ and Moon illum.
$\sigma_{300\mathrm{s}}$ is the standard deviation of residuals
binned every 5 minute, and $\beta$ is the ratio between the noise
levels due to Poisson noise and combined Poisson and red noise.}
\setlength{\tabcolsep}{5pt}
\begin{tabular}{lcccccccccccc} \hline
Occultation & Date of   & Start time & End time  &$N_{\rm obs}$ & $T_{\rm exp}$ & $T_{\rm obs}$ & Filter & Airmass & Moon & Aperture   & $\sigma_{300\mathrm{s}}$ & $\beta$ \\
& first obs &    (UT)    &   (UT)    &              & (s)           & (s)           &        &         &illum.& radii (px) & (ppm)  \\
\hline
DFOSC & 2010 05 23 & 23:03 & 03:01 &  91 & 120 & 155 & Gunn $i$ & 1.05 $\to$ 1.73 & 0.60 & 18, 24, 32  & 547 & 1.00 \\
DFOSC & 2011 05 13 & 22:59 & 02:34 & 100 &  90 & 115 & Gunn $i$ & 1.04 $\to$ 1.35 & 0.84 & 22, 32, 39  & 420 & 1.16 \\
DFOSC & 2011 05 24 & 23:09 & 03:28 &  94 & 120 & 155 & Gunn $i$ & 1.03 $\to$ 2.01 & 0.48 & 21, 27, 35  & 537 & 1.00 \\
DFOSC & 2011 06 08 & 23:23 & 03:11 &  84 & 120 & 115 & Gunn $i$ & 1.13 $\to$ 2.44 & 0.50 & 18, 24, 32  & 816 & 1.12 \\
\hline \end{tabular} \end{table*}
%
%

\section{Light-curve analysis}
\label{sec:3}
%
\begin{figure*} %
\includegraphics[width=\textwidth]{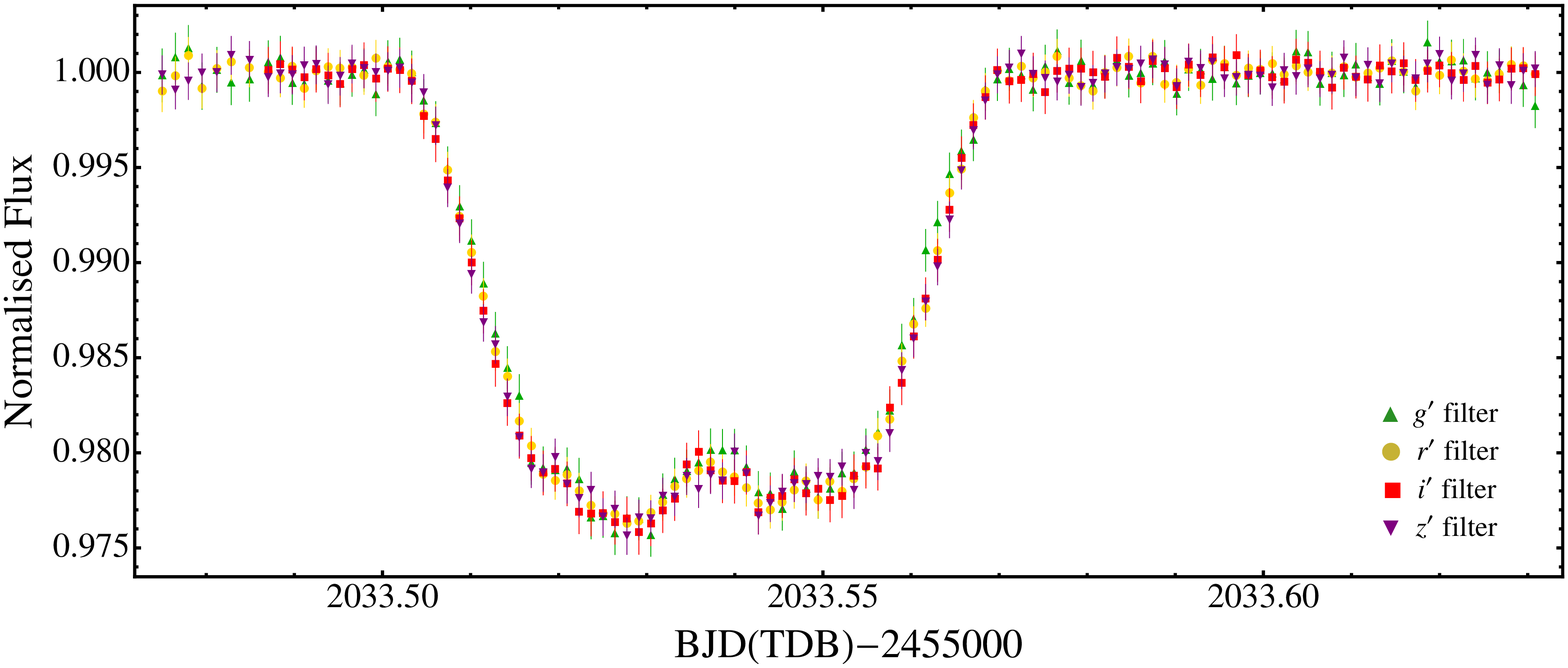}
\caption{\label{fig:05} Superimposed optical light curves of
WASP-19 obtained with GROND. Green triangles are for the data
taken in the $g^{\prime}$ band, yellow dots for $r^{\prime}$, red
squares for $i^{\prime}$ and purple inverted triangles for
$z^{\prime}$. The bump observed near the centre is due to the
occultation of a starspot by the planet.}
\end{figure*}

Fig.\,\ref{fig:02} shows the multi-band GROND light curves of one
transit. There is an anomaly near to the midpoint of the transit
which could not be removed by choosing different comparison stars.
The light curves are shown superposed in Fig.\,\ref{fig:05} to
allow comparison of the transit depth and starspot anomaly in each
passband. Following \citet{tregloan2013}, we interpret this
anomaly as the consequence of a dark starspot on the stellar
photosphere. These light curves were analysed with a dedicated
code and discussed in Sect.\,\ref{sec:3.3}. We do not see any
colour-depending variation in the transit depth similar to that
detected by \citet{lendl2012} in their simultaneous observation of
a transit of WASP-19 with two different telescopes and optical
broad-band filters.

The other WASP-19 transit light curves presented in
Sect.\,\ref{sec:2.1} were analysed using the {\sc
jktebop}\footnote{\textsc{jktebop} is written in FORTRAN77 and the
source code is available at: {\tt
www.astro.keele.ac.uk/jkt/codes/jktebop.html}} code
\citep{southworth2004}. The main fitted parameters were the
orbital inclination, $i$, the transit midpoint, $T_0$, and the sum
and ratio of the fractional radii of the star and planet,
$r_{\mathrm{A}}+r_{\mathrm{b}}$ and $k =
r_{\mathrm{b}}/r_{\mathrm{A}}$, where $r_{\mathrm{A}} =
R_{\mathrm{A}}/a$ and $r_{\mathrm{b}} = R_{\mathrm{b}}/a$, $a$ is
the orbital semimajor axis, and $R_{\mathrm{A}}$ and
$R_{\mathrm{b}}$ are the absolute radii of the star and the
planet, respectively. The occultation light curves were analysed
with a different procedure (see Sect.\,\ref{sec:6}). The orbital
eccentricity was fixed to zero \citep{anderson2011} throughout the
analysis.

\subsection{Orbital period determination}
\label{sec:3.1}

\begin{table} \begin{center}
\caption{\label{tab:03} Times of minimum light of WASP-19 and
their residuals versus the ephemeris derived in this work.}%
\tiny
\begin{tabular}{l@{\,$\pm$\,}l r r l} \hline
\multicolumn{2}{l}{Time of minimum}   & Cycle  & Residual & Reference \\
\multicolumn{2}{l}{(BJD(TDB) $-$ 2400000)} & number & (JD)    &           \\
\hline %
54775.33720 & 0.00150 &    0 &-0.000606 & \citet{hebb2010}      \\
54776.91566 & 0.00019 &    2 & 0.000178 & \citet{anderson2010}  \\
54817.14633 & 0.00021 &   53 & 0.000046 & \citet{lendl2012}     \\
55199.73343 & 0.00083 &  538 & 0.000134 & Tifner F. (TRESCA)    \\
55251.79657 & 0.00014 &  604 &-0.000113 & \citet{tregloan2013}  \\
55252.58544 & 0.00010 &  605 &-0.000082 & \citet{tregloan2013}  \\
55255.74077 & 0.00012 &  609 &-0.000109 & \citet{tregloan2013}  \\
55259.68448 & 0.00033 &  614 &-0.000595 & Colque J. (TRESCA)    \\
55273.88282 & 0.00062 &  632 &-0.001360 & Evans P.  (TRESCA)    \\
55299.12768 & 0.00055 &  664 & 0.000645 & Curtis I. (TRESCA)    \\
55334.62540 & 0.00021 &  709 & 0.000604 & This work (DFOSC)     \\
55338.56927 & 0.00023 &  714 & 0.000275 & \citet{lendl2012}     \\
55353.55659 & 0.00024 &  733 &-0.000347 & This work (DFOSC)     \\
55363.81131 & 0.00041 &  746 &-0.000539 & Milne G.  (TRESCA)    \\
55368.54285 & 0.00212 &  752 &-0.002032 & This work (DFOSC)     \\
55539.72327 & 0.00030 &  969 & 0.000279 & \citet{lendl2012}     \\
55569.69826 & 0.00036 & 1007 &-0.000620 & \citet{lendl2012}     \\
55583.10979 & 0.00089 & 1024 & 0.000643 & Curtis I. (TRESCA)    \\
55584.68693 & 0.00024 & 1026 & 0.000015 & \citet{lendl2012}     \\
55584.68684 & 0.00019 & 1026 & 0.000105 & \citet{lendl2012}     \\
55594.15188 & 0.00168 & 1038 &-0.001018 & This work (PEST)      \\
55601.25164 & 0.00071 & 1047 &-0.000813 & This work (PEST)      \\
55602.83138 & 0.00046 & 1049 & 0.001253 & \citet{lendl2012}     \\
55605.19414 & 0.00180 & 1052 &-0.002500 & This work (PEST)      \\
55606.77464 & 0.00022 & 1054 & 0.000317 & \citet{lendl2012}     \\
55607.56241 & 0.00033 & 1055 &-0.000752 & \citet{lendl2012}     \\
55622.55057 & 0.00026 & 1074 &-0.000537 & \citet{lendl2012}     \\
55624.12787 & 0.00142 & 1076 &-0.000912 & This work (PEST)      \\
55632.80612 & 0.00025 & 1087 & 0.000103 & \citet{lendl2012}     \\
55655.68222 & 0.00045 & 1116 &-0.000133 & \citet{lendl2012}     \\
55670.66976 & 0.00064 & 1135 &-0.000538 & \citet{lendl2012}     \\
55677.77038 & 0.00195 & 1144 & 0.000531 & This work (PEST)      \\
55688.81201 & 0.00333 & 1158 &-0.001592 & This work (PEST)      \\
55689.60276 & 0.00030 & 1159 & 0.000324 & This work (DFOSC)     \\
55692.75674 & 0.00255 & 1163 &-0.001054 & This work (PEST)      \\
55693.54639 & 0.00013 & 1164 &-0.000245 & This work (DFOSC)     \\
55703.79933 & 0.00411 & 1177 &-0.002214 & This work (PEST)      \\
55704.59078 & 0.00034 & 1178 & 0.000396 & This work (DFOSC)     \\
55708.53495 & 0.00015 & 1183 & 0.000374 & This work (DFOSC)     \\
55886.81234 & 0.00208 & 1409 & 0.000100 & This work (PEST)      \\
55896.27611 & 0.00210 & 1421 &-0.002195 & This work (PEST)      \\
55915.20980 & 0.00065 & 1445 &-0.000646 & This work (PEST)      \\
55919.15485 & 0.00103 & 1450 & 0.000200 & This work (PEST)      \\
55922.30966 & 0.00555 & 1454 &-0.000343 & This work (PEST)      \\
55999.61630 & 0.00007 & 1552 & 0.000057 & \citet{bean2013}      \\
56021.70374 & 0.00009 & 1580 & 0.000002 & \citet{bean2013}      \\
56029.59250 & 0.00035 & 1590 & 0.000366 & \citet{lendl2012}     \\
56033.53645 & 0.00014 & 1595 & 0.000097 & This work (GROND $g$) \\
56033.53651 & 0.00007 & 1595 & 0.000122 & This work (GROND $r$) \\
56033.53643 & 0.00007 & 1595 & 0.000180 & This work (GROND $i$) \\
56033.53652 & 0.00009 & 1595 & 0.000187 & This work (GROND $z$) \\
56063.51174 & 0.00030 & 1633 &-0.000480 & \citet{lendl2012}     \\
56334.87208 & 0.00053 & 1977 &-0.000823 & Evans P.  (TRESCA)    \\
\hline %
\end{tabular} \end{center} \end{table}

%
\begin{figure*} %
\includegraphics[width=18.0cm]{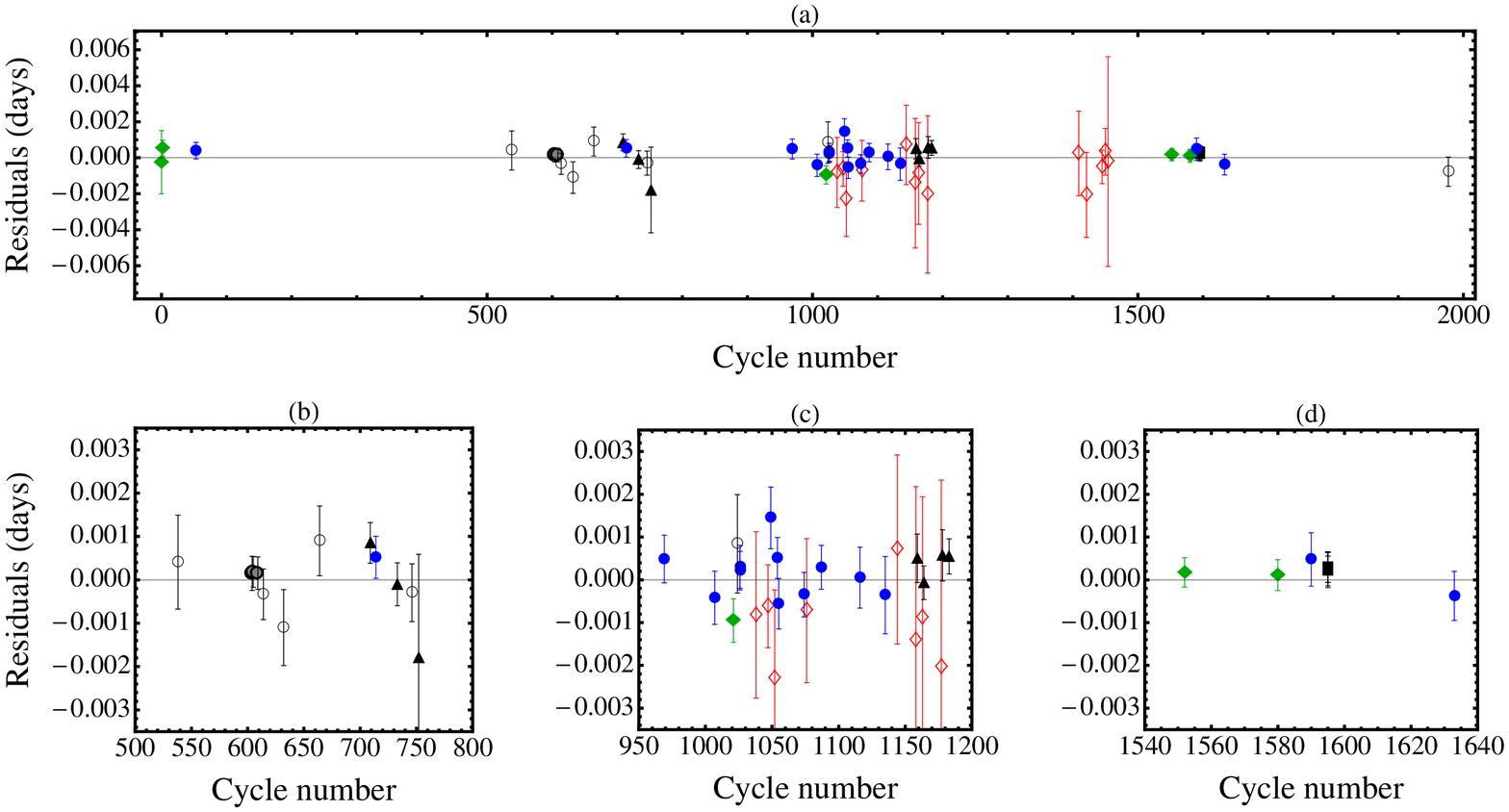}
\caption{\label{fig:minima} \emph{Panel (a)}: plot of the
residuals of the timings of mid-transit of WASP-19 versus a linear
ephemeris. The timings plotted with black triangles are from DFOSC
(this work), red open diamonds from PEST (this work), black boxes
from GROND (this work), blue circles from \citet{lendl2012}, grey
circles from \citep{tregloan2013}, black open circles from ETD,
and green diamonds show literature values (see
Table\,\ref{tab:03}.) \emph{Panels (b), (c) and (d)}: zooms in to
the best-sampled regions.}
\end{figure*}

We used our photometric data to refine the orbital period of
WASP-19\,b. The transit time for each of our data sets was
obtained by fitting them with {\sc jktebop}, and uncertainties
were estimated by Monte Carlo simulations. To these we added three
timings from high-precision light curves obtained by
\citep{tregloan2013}, 14 timings from \citet{lendl2012}, five from
literature sources and another seven measured by amateur
astronomers and available on the ETD\footnote{The Exoplanet
Transit Database (ETD) website can be found at
http://var2.astro.cz/ETD} website, for a total of 54 times of
mid-transit. The ETD light curves were included only if they had
complete coverage of the transit and a Data Quality index $\leq
3$. We recalculated the timing from \citet{dragomir2011} by
fitting this light curve with {\sc jktebop} (Fig.\,\ref{fig:03}).
All timings were placed on BJD(TDB) time system and are summarized
in Table\,\ref{tab:03}. The resulting measurements of transit
midpoints were fitted with a straight line to obtain a final
orbital ephemeris:
\begin{equation}
T_{0} = \mathrm{BJD(TDB)} 2\,454\,775.33745(35) +
0.7888396(10)\,E, \nonumber
\end{equation}
where $E$ is the number of orbital cycles after the reference
epoch, which we take to be the midpoint of the first transit
observed by \citet{hebb2010}, and quantities in brackets denote
the uncertainty in the final digit of the preceding number. The
fit has $\chi_{\nu}^2=1.98$, which is significantly greater than
unity. The uncertainties in the ephemeris given above have been
increased to account for this. The large $\chi_{\nu}^2$ implies
that the error bars in the various $T_0$ measurements are too
small. A plot of the residuals around the fit is shown in
Fig.\,\ref{fig:minima} and does not indicate any clear systematic
deviation from the predicted transit times. Based on our
experience with a similar situation in previous studies (e.g.\
\citealp{southworth2012a,southworth2012b,mancini2013a}), we
conservatively do not interpret the large $\chi_{\nu}^2$ as
indicating transit timing variations.

\subsection{Optical light-curve modelling}
\label{sec:3.2}
%
\begin{table*} %
\caption{\label{tab:04} Parameters of the fits to the light curves
of WASP-19 from the {\sc prism+gemc} (GROND data) and {\sc
jktebop} (DFOSC and PEST data) analyses. The complementary results
from \citet{tregloan2013} are labelled NTT. The final parameters
are given in bold and the parameters found by other studies are
also shown at the base of the table for comparison. Quantities
without quoted uncertainties were not given by those authors but
have been calculated from other parameters which were.
\newline {\bf Notes:}
$^{a}$The results in this row come from the analysis of eight
light curves combined by phase-binning. $^{b}$The results in this
row come from the analysis of 14 light curves combined by
phase-binning. $^{c}$The results in this row come from the
analysis of three NTT light curves performed by
\citet{tregloan2013}.} \setlength{\tabcolsep}{4pt}
\begin{tabular}{l c r@{\,$\pm$\,}l r@{\,$\pm$\,}l r@{\,$\pm$\,}l r@{\,$\pm$\,}l r@{\,$\pm$\,}l}
\hline
Source & \hspace*{-10pt}Filter & \mc{$r_{\rm A}+r_{\rm b}$} & \mc{$k$} & \mc{$i$ ($^\circ$)} & \mc{$r_{\rm A}$} & \mc{$r_{\rm b}$} \\
\hline
GROND       & $g^{\prime}$ & 0.33106 & 0.00085 & 0.14206 & 0.00038 & 78.39 & 0.42 & 0.28987 & 0.00076 & 0.04118 & 0.00014 \\
GROND       & $r^{\prime}$ & 0.33437 & 0.00374 & 0.14372 & 0.00056 & 78.37 & 0.28 & 0.29238 & 0.00332 & 0.04202 & 0.00050 \\
GROND       & $i^{\prime}$ & 0.32958 & 0.00450 & 0.14386 & 0.00080 & 78.98 & 0.36 & 0.28822 & 0.00388 & 0.04146 & 0.00059 \\
GROND       & $z^{\prime}$ & 0.32983 & 0.00410 & 0.14207 & 0.00058 & 78.95 & 0.31 & 0.28883 & 0.00360 & 0.04103 & 0.00053 \\
DFOSC$^{a}$ & $R$          & 0.32096 & 0.00843 & 0.14028 & 0.00121 & 79.40 & 0.77 & 0.28148 & 0.00717 & 0.03949 & 0.00130 \\
PEST$^{b}$  & $R$          & 0.34401 & 0.01809 & 0.14722 & 0.00270 & 77.42 & 1.31 & 0.29986 & 0.01525 & 0.04415 & 0.00287 \\
NTT$^{c}$   & $r$          & 0.33010 & 0.00190 & 0.14280 & 0.00060 & 78.94 & 0.23 & 0.28884 & 0.00165 & 0.04125 & 0.00028 \\
\hline
Final results&&{\bf0.33091}&{\bf0.00074}&{\bf0.14259}&{\bf0.00023}&{\bf78.76}&{\bf0.13}&{\bf0.28968}&{\bf0.00065}&{\bf0.04124}&{\bf0.00012}\\
\hline
\citet{tregloan2013}\hspace*{-10pt}\, & & \mc{0.3301 $\pm$ 0.0019} & \mc{0.1428 $\pm$ 0.0006} & \mc{78.94 $\pm$ 0.23} & \mc{0.2888} & \mc{0.04125} \\
\citet{lendl2012}    & & \mc{0.3193} & \mc{0.1423} & \mc{79.54 $\pm$ 0.33} & \mc{0.2796} & \mc{0.0398} \\
\citet{bean2013}     & & \mc{ } & \mc{0.0207} & \mc{78.73 $\pm$ 0.20} & \mc{ } & \mc{ } \\
\citet{abe2013}      & & \mc{ } & \mc{0.141 $\pm$ 0.001} & \mc{79.6 $\pm$ 0.3} & \mc{ } & \mc{ } \\
\hline %
\end{tabular}
\end{table*} %
%
\begin{table} %
\centering \caption{\label{tab:05} Spectroscopic properties of the
host star in WASP-19 adopted from the literature and used in the
determination of the LD coefficients and the physical properties
of the system.}
\begin{tabular}{l r@{\,$\pm$\,}l c }
\hline
Source      & \mc{WASP-19} & Ref.  \\
\hline
\Teff\ (K)             & 5460  & 90    & \citet{doyle2013}   \\
\FeH\ (dex)            & 0.14  & 0.11  & \citet{doyle2013}   \\
$\log g_{\rm A}$ (cgs) & 4.432 & 0.013 & \citet{hellier2011} \\
$K_{\rm A}$ (\ms)      & 257   & 3     & \citet{hellier2011} \\
\hline %
\end{tabular}
\end{table}

The light curves from DFOSC were individually analysed, while the
PEST ones were normalized to zero magnitude, phase-binned and
fitted with {\sc jktebop}. We used a quadratic law to model the
limb darkening (LD) of the star. For each instrument, the two LD
coefficients were selected considering the filter used and their
theoretically predicted values from \citet{claret2004}. The
atmospheric parameters of the star adopted to derive the LD
coefficients are shown in Table\,\ref{tab:05}. The linear LD
coefficients were fitted, whereas the non-linear ones were fixed
but perturbed during the process of error estimating.
Uncertainties in the fitted parameters from each solution were
calculated from 3000 Monte Carlo simulations and by a
residual-permutation algorithm \citep{southworth08}. The larger of
the two possible error bars was retained in each case. The light
curves and their best-fitting models are shown in
Fig.\,\ref{fig:01} for the DFOSC light curves and in
Fig.\,\ref{fig:03} for the PEST data. An investigation of the
residuals in the right-hand panel of Fig.\,\ref{fig:01} suggests
that they are probably contaminated by systematic noise. Another
explanation is related to the starspot activity in the photosphere
of the parent star. This possibility is discussed in Section
\ref{sec:3.3}. Finally, the DFOSC light curves were combined by
phase, binned and refitted with {\sc jktebop}. The final light
curve has a scatter around the best-fitting model of 0.58\,mmag.

The parameters of each fit with their uncertainties are reported
in Table\,\ref{tab:04}. They were combined with those coming from
the analysis of the GROND optical data (see Sect.\,\ref{sec:3.3}),
whereas the GROND NIR light curves were ignored at this point due
to their much larger scatter (Fig.\,\ref{fig:02}). The final
photometric parameters are the weighted mean of all the results
presented in Table\,\ref{tab:04}. Values obtained by other authors
are also reported for comparison.

\subsection{Starspot modelling} %
\label{sec:3.3}
%
We modelled the GROND transit light curves of WASP-19 with the
{\sc prism}\footnote{Planetary Retrospective Integrated Star-spot
Model.} and {\sc gemc}\footnote{Genetic Evolution Markov Chain.}
codes \citep{tregloan2013}. The first code models a planetary
transit over a spotted star, while the latter one is an
optimisation algorithm for finding the global best fit and
associated uncertainties. Using these codes, it is then possible
not only to determine $k$, $r_{\mathrm{A}}+r_{\mathrm{b}}$, $i$,
$T_{0}$ and the LD coefficients, but also the parameters of the
starspot. The starspot parameters are the projected longitude and
the latitude of its centre ($\theta$ and $\phi$), its angular
radius $r_{\mathrm{spot}}$ and its contrast
$\rho_{\mathrm{spot}}$, the latter being the ratio of the surface
brightness of the starspot to that of the surrounding photosphere.

Since the codes are not able to simultaneously fit the four GROND
data sets, we performed a two-step analysis. First, we modelled
all the GROND data sets separately and determined the photometric
parameters which we reported in Table\,\ref{tab:04}. These were
used to revise the physical properties of the system (see
Sect.\,\ref{sec:4}). In the second step, we combined the four
light curves into a single data set by taking the mean value at
each point from the four optical bands at that point and we fitted
the corresponding light curve. In this way we found a common value
for $T_0$, $i$, $\theta$ and $\phi$. The values of the latter two
quantities are $\theta=3.36^{\circ}\pm0.08^{\circ}$ and
$\phi=59.98^{\circ}\pm0.80^{\circ}$. Finally, we refitted each
light curve separately, this time fixing $\theta$ and $\phi$ to
the values found in the combined fit, and deriving
$r_{\mathrm{spot}}$ and $\rho_{\mathrm{spot}}$ in each band
(Table\,\ref{tab:06}).

\begin{table*} %
\centering %
\caption{\label{tab:06} Starspot parameters derived from the {\sc
prism+gemc} fitting of the optical GROND transit light curves for
a common starspot position ($\theta=3.36^{\circ}$,
$\phi=59.98^{\circ}$).}
\begin{tabular}{l c | c c c c}%
\hline %
Parameter & Symbol & $g^{\prime}$ & $r^{\prime}$ & $i^{\prime}$ & $z^{\prime}$ \\
\hline %
Starspot angular radius ($^{\circ}$)  & $r_{\mathrm{spot}}$    & $9.37  \pm 0.45 $ & $9.65  \pm 0.50 $ & $10.50 \pm 0.67 $ & $8.60  \pm 0.57 $ \\
Starspot contrast                     & $\rho_{\mathrm{spot}}$ & $0.347 \pm 0.050$ & $0.590 \pm 0.037$ & $0.638 \pm 0.020$ & $0.618 \pm 0.040$ \\
\hline %
\end{tabular}
\end{table*}

The final value for the spot angular radius comes from the
weighted mean of the results in each band and is
$r_{\mathrm{spot}}=9.46^{\circ} \pm 0.26^{\circ}$. This translates
to a radius of $116\,900 \pm 3220$\,km, which is equivalent to a
size of $2.7\%$ of the stellar disc and is $\sim 68\,460$\,km
smaller than the starspot found by \citet{tregloan2013}. The sizes
of starspots are most often estimated through Doppler-imaging
techniques (e.g.\ \citealp{cameron1992,vogt1999}), which are
unable to measure starspots with a size less than $0.1\%$ of a
stellar hemisphere \citep{strassmeier2009}. However, our
measurement is similar to those found for other G-type stars and
is in good agreement with the sizes of common sunspots.

\begin{figure*}
\includegraphics[width=16.0cm]{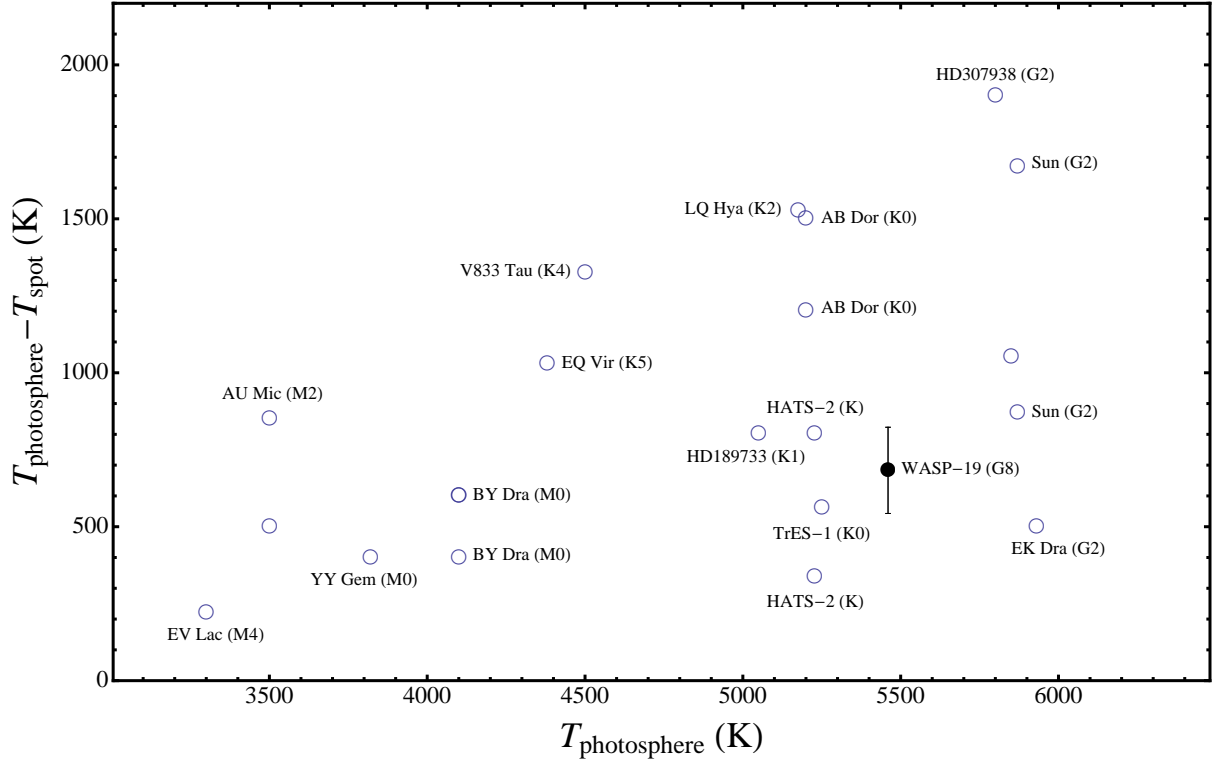}
\caption{Spot temperature contrast with respect to the
photospheric temperature in several dwarf stars taken from
\citet{berdyugina2005}. The name of the star and its spectral type
are also reported for most of them. The values for TrES-1,
HD189733 and HATS-2 are taken from \citet{rabus2009},
\citet{sing2011} and \citet{mohler2013},
respectively. The black dot refers to WASP-19 (this work). Note that some stars appear twice.} %
\label{fig:07}
\end{figure*}

It is also interesting to study how starspot contrast changes with
passband. Starspots are expected to be darker in the ultraviolet
(UV) than in the IR. In our case, as can be seen from
Table\,\ref{tab:06}, moving from $g^{\prime}$ to $z^{\prime}$, the
starspot becomes brighter (note that the values for $i^{\prime}$
and $z^{\prime}$ are well within $1\sigma$). Modelling both the
photosphere and the starspot as black bodies
\citep{rabus2009,sanchis2011} and using Eq.\,1 of
\citet{silva2003} and $T_{\mathrm{eff}}=5460\pm90$
\citep{doyle2013}, we found the temperature of the starspot in
each band: $T_{\mathrm{spot},g}= 4595 \pm 118$\,K,
$T_{\mathrm{spot},r}= 4864 \pm 96$\,K, $T_{\mathrm{spot},i}= 4842
\pm 81$\,K and $T_{\mathrm{spot},z}= 4698 \pm 112$\,K. The
weighted mean is $T_{\mathrm{spot}}=4777 \pm 50$\,K, and is
consistent with what has been observed for other stars (see
Fig.\,\ref{fig:07}) and for the case of the TrES-1
\citep{rabus2009}, HD189733 \citep{sing2011}, and HATS-2
\citep{mohler2013}.
%
%
\begin{figure}
\includegraphics[width=9.0cm]{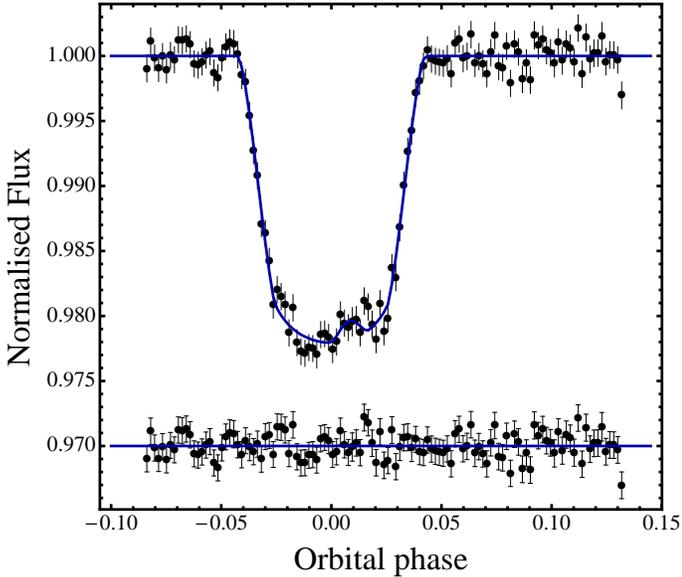}
\caption{Transit light curve of WASP-19 observed with the Danish
1.54\,m telescope on 2011/05/11. The best-fitting line has been
calculated by using {\sc prism+gemc} codes and is compatible with the presence of a starspot.} %
\label{fig:08}
\end{figure}

Finally, we investigated whether the systematic residuals from
fitting the transit DFOSC light curves can be attributed to
starspots. We selected one of the best DFOSC light curves, that of
2011/05/11, and refitted it using {\sc prism+gemc}. The fit
includes a starspot just after mid-transit with starspot
parameters in reasonable agreement with that of the starspot
detected in the GROND data (this work) and in the NTT data
\citep{tregloan2013}. The best fit is visible in
Fig.\,\ref{fig:08} and the fitted parameters are shown in
Table\,\ref{tab:07}, together with those obtained without
considering the starspot. The presence of the starspot cannot be
confirmed because the amplitude of the anomaly is similar to the
scatter of the data points.
%
\begin{table*} %
\caption{\label{tab:07} Parameters of the fit to the light curve
of WASP-19, observed with the Danish 1.54\,m telescope on
2011/05/11, from the {\sc jktebop} and {\sc prism+gemc} analysis.}
\setlength{\tabcolsep}{4pt}
\begin{tabular}{l r r@{\,$\pm$\,}l r@{\,$\pm$\,}l r@{\,$\pm$\,}l r@{\,$\pm$\,}l r@{\,$\pm$\,}l r@{\,$\pm$\,}l r@{\,$\pm$\,}l}
\hline
Source & code       & \mc{$r_{\rm A}+r_{\rm b}$} & \mc{$k$} & \mc{$i$ ($^\circ$)} & \mc{$\theta$} & \mc{$\phi$} & \mc{$r_{\rm spot}$} & \mc{$\rho_{\rm spot}$}  \\
\hline
DFOSC & {\sc jktebop}    & 0.3194 & 0.0060 & 0.13774 & 0.00097 & 79.38 & 0.95  \\
DFOSC & {\sc prism+gemc} & 0.3250 & 0.0087 & 0.13817 & 0.00107 & 79.09 & 0.67 & 11.35 & 0.89 & 59.11 & 5.49 & 6.71 & 0.72 & 0.66 & 0.06 \\
\hline %
\end{tabular}
\end{table*} %

\section{Physical properties}
\label{sec:4}

We measured the physical properties of the WASP-19 system using
the \emph{Homogeneous Studies} methodology (see
\citealt{southworth12c} and references therein). Briefly, we used
the photometric parameters $r_{\rm A}$, $r_{\rm b}$ and $i$
(Sect.\,\ref{sec:3}), published spectroscopic results (see
Table\,\ref{tab:05}), and theoretical stellar models to estimate
the properties of the system. This was done for a range of ages to
find the best-fitting evolutionary age of the star, and for five
different sets of theoretical models \citep{southworth10} to
determine the systematic error stemming from their use. The random
errors in all input parameters were propagated to all output
parameters using a perturbation analysis. We also used a
model-independent method to estimate the physical parameters of
the systems, via a calibration based on detached eclipsing
binaries relating the host star's density to its radius
\citep{enoch2010}. Our implementation relies on the calibration
coefficients for stars less than 3\Msun\ calculated by
\citet{southworth11}.

The individual solutions for each stellar model and for the
empirical calibration method can be found in Table\,\ref{tab:08}.
The final set of physical properties was obtained by taking the
unweighted mean of the sets of values obtained using stellar
models, and are reported in Table\,\ref{tab:09}. The accuracy of
our final estimation of most of the parameters is better than
those obtained by other authors. In particular, the radius of the
planet was measured with a precision better than $2\%$.

\begin{table*} \caption{\label{tab:08} Derived physical properties of the WASP-19 planetary system.}
\setlength{\tabcolsep}{4pt}
\begin{tabular}{l r@{\,$\pm$\,}l r@{\,$\pm$\,}l r@{\,$\pm$\,}l r@{\,$\pm$\,}l r@{\,$\pm$\,}l r@{\,$\pm$\,}l}
\hline
\ & \mc{This work} & \mc{This work} & \mc{This work} & \mc{This work} & \mc{This work} & \mc{This work} \\
\ & \mc{(dEB constraint)} & \mc{({\sf Claret} models)} & \mc{({\sf Y$^2$} models)} & \mc{({\sf Teramo} models)} & \mc{({\sf VRSS} models)} & \mc{({\sf DSEP} models)} \\
\hline
$K_{\rm b}$     (\kms) & 230.3     &   6.1      & 222.4     &   3.1      & 222.7     &   3.2      & 219.6     &   2.4      & 218.8     &   2.2      & 220.4     &   2.2      \\
$M_{\rm A}$    (\Msun) & 1.060     & 0.085      & 0.955     & 0.040      & 0.960     & 0.041      & 0.920     & 0.030      & 0.910     & 0.027      & 0.929     & 0.028      \\
$R_{\rm A}$    (\Rsun) & 1.062     & 0.028      & 1.025     & 0.014      & 1.027     & 0.015      & 1.013     & 0.011      & 1.009     & 0.011      & 1.016     & 0.010      \\
$\log g_{\rm A}$ (cgs) & 4.4118    & 0.0117     & 4.3966    & 0.0064     & 4.3973    & 0.0066     & 4.3912    & 0.0052     & 4.3896    & 0.0046     & 4.3927    & 0.0048     \\[2pt]
$M_{\rm b}$    (\Mjup) & 1.239     & 0.067      & 1.156     & 0.035      & 1.160     & 0.036      & 1.128     & 0.028      & 1.119     & 0.027      & 1.135     & 0.026      \\
$R_{\rm b}$    (\Rjup) & 1.471     & 0.040      & 1.421     & 0.021      & 1.423     & 0.021      & 1.403     & 0.016      & 1.398     & 0.015      & 1.408     & 0.015      \\
$\rho_{\rm b}$ (\pjup) & 0.3643    & 0.0115     & 0.3772    & 0.0083     & 0.3766    & 0.0084     & 0.3819    & 0.0078     & 0.3834    & 0.0078     & 0.3806    & 0.0075     \\
\safronov\             & 0.02706   & 0.00079    & 0.02802   & 0.00052    & 0.02798   & 0.00053    & 0.02838   & 0.00047    & 0.02848   & 0.00047    & 0.02828   & 0.00045    \\[2pt]
$a$               (AU) & 0.01704   & 0.00045    & 0.01646   & 0.00023    & 0.01649   & 0.00024    & 0.01626   & 0.00018    & 0.01620   & 0.00016    & 0.01631   & 0.00016    \\
Age              (Gyr) &       \mc{ }        & \erc{ 9.8}{ 2.6}{ 1.9} & \erc{ 8.0}{ 2.1}{ 1.5} & \erc{11.6}{ 2.1}{ 2.0} & \erc{11.7}{ 0.9}{ 3.0} & \erc{10.0}{ 2.4}{ 1.2} \\
\hline
\end{tabular}
\end{table*}

\begin{table*} %
\caption{\label{tab:09} Final physical properties of the WASP-19
system, compared with results from the literature. Separate
statistical and systematic errorbars are given for the results
from the current work.}
\setlength{\tabcolsep}{4pt}
\begin{tabular}{l r@{\,$\pm$\,}c@{\,$\pm$\,}l r@{\,$\pm$\,}l r@{\,$\pm$\,}l r@{\,$\pm$\,}l r@{\,$\pm$\,}l r@{\,$\pm$\,}l r@{\,$\pm$\,}l r@{\,$\pm$\,}l}
\hline
\ & \mcc{This work} & \mc{\citet{hebb2010}} & \mc{\citet{hellier2011}} & \mc{\citet{tregloan2013}} & \mc{\citet{lendl2012}} \\
\hline
$M_{\rm A}$    (\Msun) & 0.935    & 0.033    & 0.025      & \erc{0.96}{0.09}{0.10}       & \mc{0.97 $\pm$ 0.02}       & \mc{0.904 $\pm$ 0.040}     & \erc{0.968}{0.084}{0.079}  \\
$R_{\rm A}$    (\Rsun) & 1.018    & 0.012    & 0.009      & \erc{0.94}{0.04}{0.04}       & \mc{0.99 $\pm$ 0.02}       & \mc{1.004 $\pm$ 0.016}     & \mc{1.004 $\pm$ 0.016}     \\
$\log g_{\rm A}$ (cgs) & 4.3932   & 0.0054   & 0.0039     & \erc{4.47}{0.03}{0.03}       & \mc{4.432 $\pm$ 0.013}     & \mc{4.391 $\pm$ 0.008}     & \mc{ }                     \\
$\rho_{\rm A}$ (\psun) & \mcc{$0.8853 \pm 0.0060$}        & \erc{1.13}{0.09}{0.09}       & \erc{0.993}{0.047}{0.042}  & \mc{0.893 $\pm$.015}       & \erc{0.983}{0.039}{0.036}  \\[2pt]
$M_{\rm b}$    (\Mjup) & 1.139    & 0.030    & 0.020      & \erc{1.15}{0.08}{0.08}       & \mc{1.168 $\pm$ 0.023}     & \mc{1.114 $\pm$ 0.036}     & \mc{1.165 $\pm$ 0.068}     \\
$R_{\rm b}$    (\Rjup) & 1.410    & 0.017    & 0.013      & \erc{1.31}{0.06}{0.06}       & \mc{1.386 $\pm$ 0.032}     & \mc{1.395 $\pm$ 0.023}     & \mc{1.376 $\pm$ 0.046}     \\
$g_{\rm b}$     (\mss) & \mcc{$14.21 \pm  0.18$}          & \erc{15.5}{1.1}{1.1}         & \mc{13.90 $\pm$ 0.68}      & \mc{14.19 $\pm$ 0.26}      & \mc{15.28 $\pm$ 0.053}     \\
$\rho_{\rm b}$ (\pjup) & 0.3800   & 0.0071   & 0.0034     & \erc{0.51}{0.06}{0.05}       & \mc{0.438 $\pm$ 0.028}     & \mc{0.384 $\pm$ 0.011}     & \erc{0.447}{0.027}{0.025}  \\[2pt]
\Teq\              (K) & \mcc{$2077 \pm   34$}            & \erc{2009}{26}{26}           & \mc{2050 $\pm$ 40}         & \mc{2067 $\pm$ 23}         & \mc{2058 $\pm$ 40}         \\
\safronov\             & 0.02823  & 0.00048  & 0.00025    & \mc{ }                       & \mc{ }                     & \mc{0.02852 $\pm$ 0.00057} & \mc{ }                     \\
$a$               (AU) & 0.01634  & 0.00019  & 0.00015    & \erc{0.0165}{0.0005}{0.0006} & \mc{0.01655 $\pm$ 0.00013} & \mc{0.01616 $\pm$ 0.00024} & \mc{0.01653 $\pm$ 0.00046} \\
Age              (Gyr) & \ermcc{10.2}{2.6}{3.1}{1.4}{2.2} & \erc{5.5}{9.0}{4.5}          & \mc{ }                     & \erc{11.5}{2.7}{2.3}       & \mc{ }                     \\
\hline \end{tabular} \end{table*}

\section{Variation of planetary radius with wavelength}
\label{sec:5}
An interesting alternative to transmission spectroscopy in probing
the atmospheres of TEPs is to study their transits with
simultaneous photometry at different wavelengths. This strategy
allows the radius of a TEP to be measured in multiple passbands
and is not affected by temporal variability, for example
starspots. The aim is to detect variations attributable to changes
in opacity at different wavelengths
\citep[e.g.][]{demooij2012,southworth2012b,mancini2013a,mancini2013b,fukui2013,nikolov2013,copperwheat2013}.

\citet{fortney2008} suggested the classification of hot Jupiters
in two classes, pM and pL, depending to the incident stellar flux
and the expected amount of absorbing substances, such as gaseous
titanium oxide (TiO) and vanadium oxide (VO), in their
atmospheres. Receiving from its parent star an incident flux of
$3.64 \times 10^{9}$\,erg\,s$^{-1}$\,cm$^{-2}$ \citep{hebb2010},
WASP-19\,b would belong to the pM class of planets. Its atmosphere
should be rich in oxidized elements, which should cause a
variation of its radius by $\sim 3\%$ between the wavelength
ranges $350$--$400$ nm and $500$--$700$ nm. Such a variation would
be directly measurable using GROND.

More recently \citet{madhusudhan2012} proposed a new
classification scheme for hydrogen-dominated exoplanetary
atmospheres, which is based on both irradiation and the
carbon-to-oxygen (C/O) ratio. The author distinguished four
classes of atmosphere (O1, O2, C1 and C2) in this 2D space, each
having distinct chemical, thermal, and spectral properties.
According to this scheme and to the temperature and incident flux
of WASP-19\,b, it should be a C2 planet, which means that it has a
C/O ratio $\geqslant 1$, even if the O2 (C/O $<$ 1) classification
is not completely ruled out and was recently supported by
\citet{lendl2012}.

\begin{figure}
\includegraphics[width=\columnwidth]{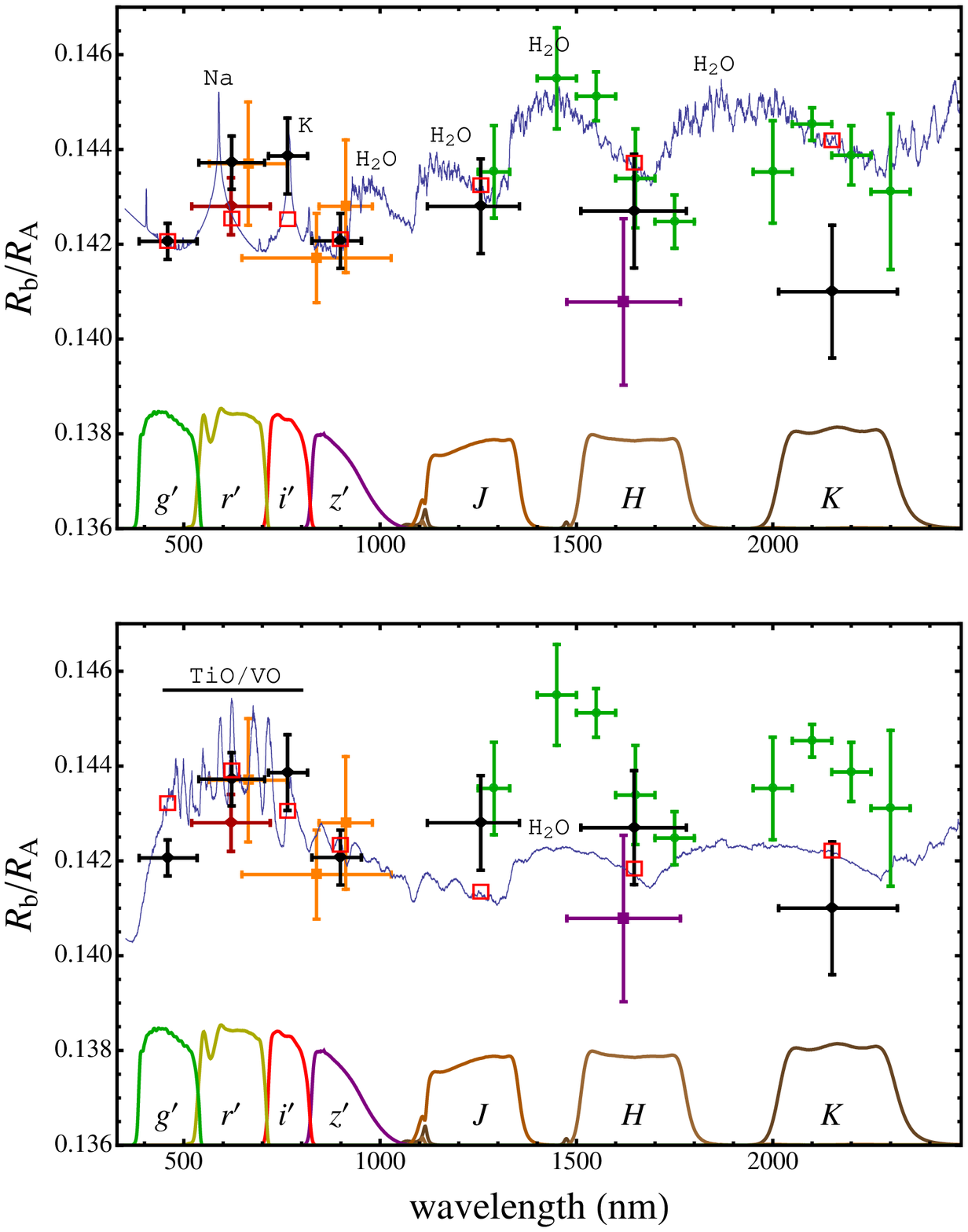}
\caption{\label{fig:09} Variation of the planetary radius, in
terms of planet/star radius ratio, with wavelength. Black diamonds
are from GROND (this work), red diamond from \citet{tregloan2013},
orange boxes from \citet{lendl2012}, purple boxes from
\citet{anderson2010}, green points from \citet{bean2013}. The
vertical bars represent the errors in the measurements and the
horizontal bars show the FWHM transmission of the passbands used.
The observational points are compared with two models. These use
profiles which are intermediate between planet-wide and day-side.
Synthetic spectra in the top panel do not include TiO and VO
opacity, while spectra in the bottom do, based on equilibrium
chemistry. Red open boxes indicate the predicted values for each
of the two models integrated over the passbands of the
observations. Transmission curves of the GROND filters are shown
at the bottom of each panel. Prominent absorption features are
labelled.}
\end{figure}

As an additional possibility offered by the GROND data, we
investigated the variations of the radius of WASP-19\,b in the
wavelength ranges accessible to the instrument and compared our
results with those available in the literature. For the optical
passbands, we used the values of the planet/star radius reported
in Table\,\ref{tab:04}. The precision of the data in the NIR bands
is significantly worse than in the optical bands (see
Fig.\,\ref{fig:02}), and the radii derived from these data are
quite uncertain. It is therefore not possible to fit for the
starspot parameters in these bands, despite the spot having a
significant effect on the measured planetary radius. A simple fit
of the NIR light curve with {\sc jktebop} returned transit depths
which are much lower than those measured in the optical cases.
This is in agreement with \citet{ballerini2012}, who showed that
the presence of starspots affects the inclination and
$a/R_{\mathrm{A}}$ values by up to 3$\sigma$. We therefore
analysed the $J$, $H$ and $K$ data sets with {\sc prism+gemc},
constraining the system and the starspot parameters (size and
position) to the values determined for the optical case. Since the
amplitude of the anomaly is expected to decrease continually
through the optical and NIR ranges, we also constrained the
fitting code to explore values of the starspot contrast $>0.7$.
With this procedure we were able to fit the NIR light curves
obtaining quite realistic transit depths. These are exhibited in
Fig.\,\ref{fig:09} together with the optical ones reported in
Table\,\ref{tab:04}, and those measured by \citet{lendl2012},
\citet{bean2013}, \citet{anderson2010} and \citet{tregloan2013}.
The vertical bars represent the relative errors in the
measurements and the horizontal bars show the FWHM transmission of
the passbands used. The GROND filter transmission curves are also
plotted at the base of each panel. A variation of the radius of
WASP-19\,b along the five passbands is quite noticeable in
Fig.\,\ref{fig:09}.

We examined this variation with the aid of 1D model atmospheres
developed in \citet{fortney2005,fortney2008,fortney2010}. By using
the chemical equilibrium abundances of \citet{Lodders02} at solar
metallicity, and the opacity database described in \citet{fr08},
we derived pressure--temperature (PT) profiles which are
intermediate between planet-wide and day-side; they are profiles
that assume $20\%$ of energy is lost to the night side (for a
planet-wide profile this number is $50\%$, whereas for a dayside
this number is set to $0\%$). We examined both the cases in which
the opacity of TiO and VO molecules is excluded or included
(Fig.\,\ref{fig:09} top and bottom panels, respectively). Coloured
open boxes indicate the predicted values for these models
integrated over the passbands of the GROND observations. Models
without TiO and VO have optical transmission spectra that are
dominated by Rayleigh scattering in the blue, and
pressure-broadened neutral atomic lines of sodium (Na) and
potassium (K) at 589\,nm and 770\,nm, respectively. Models with
TiO and VO absorption show significant optical absorption that
broadly peaks around 700\,nm, with a sharp fall-off in the blue,
and a shallower fall-off in the red.

Comparing the two panels of Fig.\,\ref{fig:09}, it is readily
apparent that the model that gives the best match to all the
observational data is the one without TiO and VO opacity in the
atmosphere of WASP-19\,b. In fact, even though both models match
the optical data quite well, the presence of strong absorbers in
the planetary atmosphere is not in agreement with the
observational results in the NIR wavelengths. Only the model in
which we neglected the presence of the two oxidized elements
succeeds to explain the optical and NIR data at the same time. In
particular the absorption features between 1200 and 1700\,nm are
quite well recovered.

\section{Secondary-eclipse analysis}
\label{sec:6}
%
We fitted the occultation light curves with a model composed of
two components:
\begin{equation}
F(\mathrm{mod})=E(T_{\mathrm{mid}},F_{\mathrm{p}}/F_*)B(x,y,t,z).
\end{equation}
The first component $E(T_{\rm{mid}},F_{\rm{p}}/F_*)$ is the real
occultation signal, which utilizes the theoretical formulae of
\citet{mandel2002} for a uniform source:
\begin{equation}
E(T_{\rm{mid}},F_{\rm{p}}/F_*)=1-\frac{\lambda_e}{1+1/(F_{\rm{p}}/F_*)},
\end{equation}
where $\lambda_e$ refers to their Equation 1, and mid-eclipse time
$T_{\rm{mid}}$ and planet-to-star flux ratio $F_{\rm{p}}/F_*$ are
set as free parameters. The second component is the baseline
correction function $B(x,y,t,z)$, which is used to correct the
instrumental systematics, consisting of the star position on the
detector ($x$, $y$), time ($t$) and airmass ($z$) in various
combinations. We find the best-fit parameters by minimizing
$\chi^2$:
\begin{equation}
\chi^2=\sum\limits_{i=1}^{N}\frac{[F_i(\mathrm{obs}) -
F_i(\mathrm{mod})]^2}{\sigma_{F,i}^2(\mathrm{obs})}.
\end{equation}
We selected the best baseline model for each night by comparing
the Bayesian Information Criterion \citep[BICs,][]{Schwarz1978}
among different models: $BIC=\chi^2+k\log(N)$.

We employed the Markov Chain Monte Carlo (MCMC) technique with
Metropolis-Hastings algorithm and Gibbs sampling to determine the
posterior probability distribution function (PDF) for each
parameter. At each step, we first divided the theoretical model
$E(T_{\rm{mid}},F_{\rm{p}}/F_*)$ out, and then solved the baseline
function coefficients using singular value decomposition
\citep[SVD,][]{press1992}. This ensures the propagation of
uncertainties from the chosen baseline function to the final PDFs.
A detailed description of our MCMC process can be found in Chen et
al.\ (2013, in preparation).

\begin{figure}
\includegraphics[width=\columnwidth]{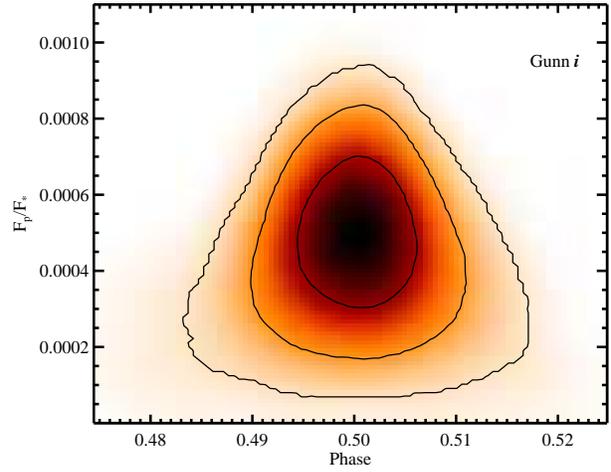}
\caption{Correlation map between mid-eclipse time and flux ratio
from the MCMC analysis. The mid-eclipse times have been converted
to phases for display purpose. The contour levels indicate
the 68.3\% (1$\sigma$), 95.4\% (2$\sigma$) and 99.7\% (3$\sigma$) confidence levels, respectively.} %
\label{fig:10}
\end{figure}

We fitted the light curves of four nights simultaneously. In this
joint fit, they shared the same $T_{\rm{mid}}$ (phase-folded) and
$F_{\rm{p}}/F_*$, but they were allowed to have different
baselines and coefficients. The planetary-system parameters were
obtained from Tables \ref{tab:05} and \ref{tab:09}. The period and
mid-transit times were those estimated in Sect.\,\ref{sec:3.1}. We
first ran five chains of 10$^5$ links to find the red noise
re-scaling factor $\beta$ and enlarged the flux uncertainties by
this factor. Then we ran another five chains of 10$^6$ links to
find the final PDFs. The results are listed in
Table\,\ref{tab:10}, while the coefficients of chosen baseline
function are listed in Table\,\ref{tab:11}. In Fig.\,\ref{fig:04}
we show the four raw light curves, as well as the
baseline-corrected phase-folded light curves. Fig.\,\ref{fig:10}
shows the joint PDF between $T_{\rm{mid}}$ and $F_{\rm{p}}/F_*$.

\begin{table}
\caption{\label{tab:10} Results of the MCMC analysis of the
secondary eclipses of WASP-19\,b.
\newline {\bf Notes:}
$^{a}$Light travel time ($\sim$16.5 s) in the system has been
corrected.}
\centering
\begin{tabular}{lcc}
\hline
Parameter & Units & Value\\
\hline
     $T_{\rm{mid,occ}}$ & BJD$_{\rm{TDB}}$ & 2455721.5508 $^{+0.0028}_{-0.0031}$\\
     $\phi_{\rm{mid,occ}}$$^{a}$  & ... & 0.5002 $^{+0.0036}_{-0.0039}$\\
     $T_{\rm{offset}}$$^{a}$  & minutes & 0.52 $^{+4.0}_{-4.5}$\\
     $F_p/F_*$ & \% & 0.048 $^{+0.013}_{-0.013}$\\
     $T_B$ & K & 2544 $^{+96}_{-110}$\\
     \hline
     \end{tabular}
\end{table}
%
\begin{table*}
\caption{\label{tab:11} Coefficients for adopted baseline
function.
\newline {\bf Note:}
$^{a} B=c_0+c_1z$; $^{b} B=c_0+c_1x$; $^{c} B=c_0+c_1x+c_2t+c_3z$;
$^{d} B=c_0+c_1x+c_2y+c_3xy+c_4y^2+c_5t$. }
     \centering
     \begin{tabular}{ccccc}
     \hline
         Coeff. & 2010-05-24$^{a}$ & 2011-05-14$^{b}$
                & 2011-05-25$^{c}$ & 2011-06-09$^{d}$\\
     \hline
        $c_{0}$ & 1.000199 $^{+0.000045}_{-0.000046}$  & 1.000364 $^{+0.000060}_{-0.000061}$
                       & 1.000117 $^{+0.000079}_{-0.000080}$
                       & 0.999764 $^{+0.000076}_{-0.000077}$\\
        $c_{1}$ & -0.002373 $^{+0.000089}_{-0.000091}$  & -0.001064 $^{+0.000098}_{-0.000097}$
                       & -0.001091 $^{+0.000032}_{-0.000032}$
                       & -0.00193 $^{+0.00019}_{-0.00018}$\\
        $c_{2}$ & ...  & ... & 0.0456 $^{+0.0039}_{-0.0039}$
                       & 0.001649 $^{+0.000071}_{-0.000070}$\\
        $c_{3}$ & ...  & ... & -0.00315 $^{+0.00059}_{-0.00059}$
                       & 0.00216 $^{+0.00022}_{-0.00022}$\\
        $c_{4}$ & ...  & ... & ...
                       & 0.001203 $^{+0.000039}_{-0.000037}$\\
        $c_{5}$ & ...  & ... & ...
                       & 0.0357 $^{+0.0013}_{-0.0014}$\\
     \hline
     \end{tabular}
\end{table*}
%

\begin{figure*}%
\includegraphics[width=18.0cm]{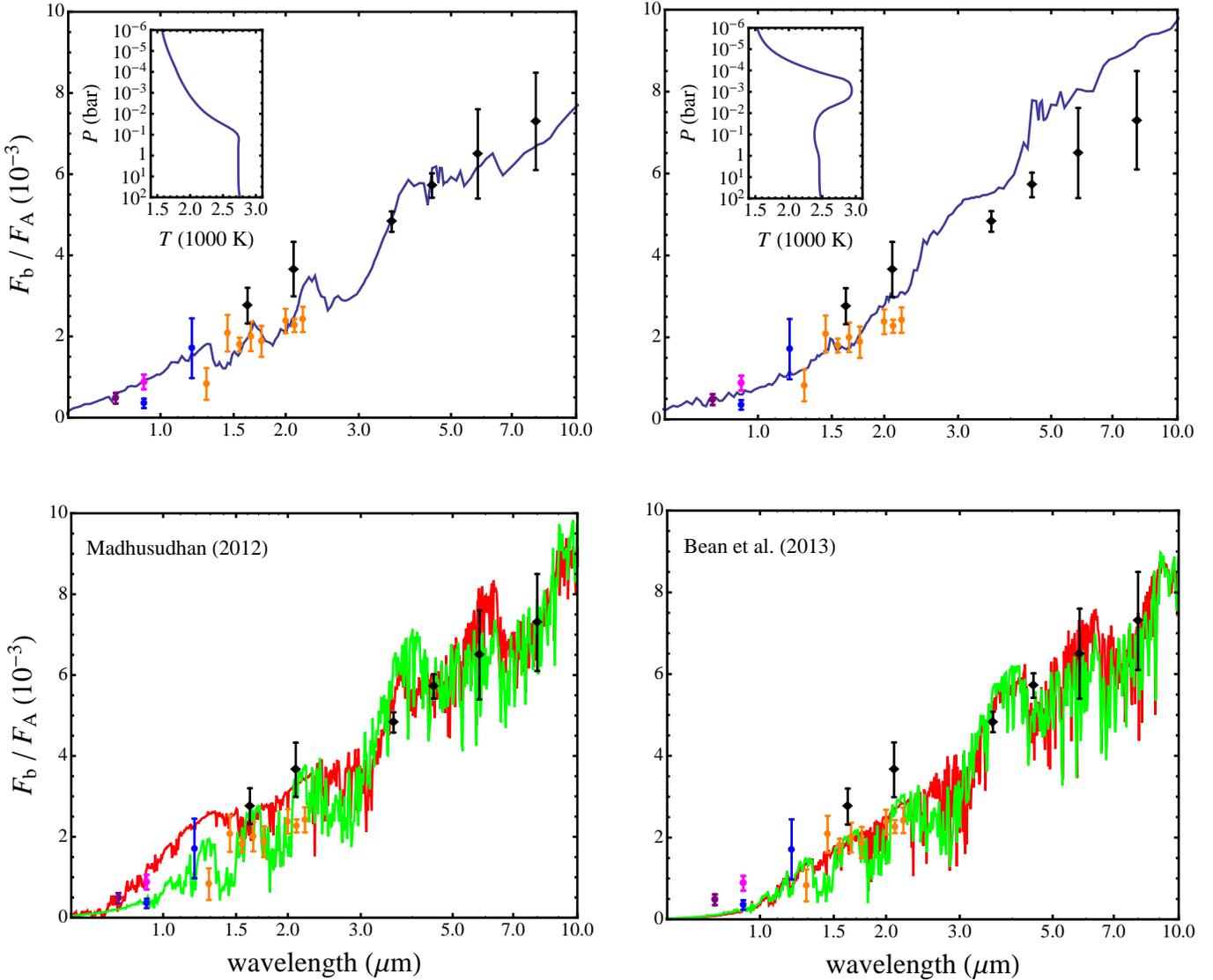}
\caption{\label{fig:11} Dayside emission spectrum of WASP-19\,b in
terms of the planet-to-star flux ratio. Our Gunn $i'$ band
detection is plotted with a purple diamond with error bars. Blue
dots are previous detections in $z'$ band and a 1.19\,{\micron}
narrow band \citep{lendl2012}. Another detection in $z'$ band is
represented by a pink dot \citep{burton2012}. Orange dots are
measurements obtained in NIR bands by \citet{bean2013}. Black
diamonds are measurements realized by \citet{anderson2010} in the
$H$ band, \citet{gibson2010} in the 2.09\,{\micron} narrow band
and \citet{anderson2011} in the four \emph{Spitzer} IRAC bands.
Different model spectra are shown in the four panels. \emph{Top
left panel}: a model atmosphere with no temperature inversion and
$\Teff=2525$\,K. \emph{Top right panel}: a model with a
temperature inversion and the same \Teff. The insets of these two
panels show the corresponding P-T profiles for the models.
\emph{Bottom left panel}: models from \citet{madhusudhan2012}. The
green line represents an oxygen-dominated (CO $=$ 0.4) model while
the red line accounts for a carbon-dominated (C/O $=$ 1.1) model.
\emph{Bottom right panel}: models taken from \citet{bean2013}. The
green and red lines are as for the bottom left panel, but with a
slightly different composition, C/O $=$ 0.5 and 1.0,
respectively.}
\end{figure*}

We have found a flux ratio of $0.048 \pm 0.013 \%$ in the Gunn $i$
band, which corresponds to a brightness temperature of
$T_{\mathrm{B}}=2544^{+96}_{-110}$\,K. In the calculation of
brightness temperature, we assumed blackbody emission for the
planet and adopted the Kurucz stellar template model for the host
star. The latter was interpolated using the \Teff\ and \FeH\ from
\citep{doyle2013}, and the $\log g_*$ obtained from this work.
This brightness temperature indicates that the day-night energy
redistribution efficiency is low ($\la 25\%$). As a matter of
fact, it is consistent with that derived from $z'$-band
\citep{burton2012}, 1.19\,\micron\ narrow band \citep{lendl2012},
NIR bands \citep{anderson2010}, and 2.09\,\micron\ narrow band
\citep{gibson2010}, see Fig.\,\ref{fig:11}. The only exception is
the other $z'$-band detection obtained by \citet{lendl2012}. It is
also important to note that the ground-based NIR spectroscopic
study on the occultations of WASP-19\,b by \citet{bean2013} shows
that it is cooler ($\sim$2250\,K) than previously determined by
\citet{anderson2010} and \citet{burton2012}.

Fig.\,\ref{fig:11} shows a comparison among different dayside
emission model spectra used to interpret the experimental data.
The top panels are models based on the methods of
\citet{fortney2008}, while bottom panels are from the literature.
In particular, the top right panel shows a model with TiO in the
atmosphere of WASP-19\,b, while the model in the top left panel
has no TiO. The insets in the two top panels show the
corresponding P-T profiles. Both models have $\Teff=2525$\,K, a
high value that is attained by allowing no loss of absorbed energy
to the night side. Therefore both of these profiles are modestly
warmer than those used for modelling the transmission spectrum.
The bottom panels of Fig.\,\ref{fig:11} show other models found in
the literature. Specifically, the \emph{bottom-left} panel shows
in green an O-dominated (C/O $=$ 0.4) model, and in red a
carbon-dominated (C/O $=$ 1.1) model \citep{madhusudhan2012}.
Quite similar models (C/O $=$ 0.5 and 1.0) are plotted in the
bottom right panel \citep{bean2013}.

From an inspection of the four panels, it seems that none of the
models are able to explain all the data simultaneously,
particularly in the NIR where there is tension between data sets.
This is not necessarily surprising since they were taken by
different instruments at different times. Since WASP-19\,A is an
active star, this indicates that it is important to do long-term
monitoring of the stellar activity. Nonetheless, whilst the model
with TiO (top-right panel of Fig.\,\ref{fig:11}) and those from
the literature (bottom panels) seem to either fit in the Spitzer
IRAC bands, or in the NIR, but not both, the model with no
temperature inversion and $\Teff=2525$\,K (top left panel) is the
one which fits well all the IRAC data, the \citet{bean2013} data
and all data blueward of 1.2\,$\mu$m, with the only exception of
that measured by \citet{lendl2012} at $0.9$\,$\mu$m. We therefore
tentatively favour an atmosphere model that lacks a temperature
inversion and has a C/O ratio that is not dramatically different
from the solar value.  A consistent picture from transit and
secondary eclipse data is that of a hot dayside atmosphere that
lacks at strong optical absorber to drive a temperature inversion.

%

\section{Summary and conclusions}
\label{sec:7}
We have reported new broad-band photometric observations of 23
transits and four occultations in the WASP-19 planetary system. It
was monitored eight times with the 1.54\,m Danish telescope and 14
times with the PEST telescope in 2011 and 2012. The transit of
2012/04/15 was obtained with GROND in seven optical-IR passbands
simultaneously, covering the wavelength range 370--2350\,nm. The
use of telescope defocussing resulted in light curves with
scatters around the best-fitting model as low as 0.52\,mmag. The
GROND data show an anomaly due to the occultation of a starspot by
the planet.

We modelled the Danish and PEST light curves using the {\sc
jktebop} code. The GROND light curves were fitted using the {\sc
prism+gemc} codes as these were designed to model starspot
anomalies. Our principal results are as follows.

\begin{itemize}
\item [$\bullet$] The position and the angular dimensions of the starspot in the
GROND optical light curves are in agreement with the starspot
found by \citet{tregloan2013} in a transit event observed in 2010
February.
\item [$\bullet$] We detected a variation of the starspot contrast as a function of
the wavelength, as expected due to the different \Teff s of the
spot and the surrounding photosphere.
\item [$\bullet$] The multi-colour data allowed the temperature of the starspot to
be constrained to be $T_{\mathrm{spot}} = 4777 \pm 50$\,K, a value
which is in agreement with those found for other similar stars
(Fig.\,\ref{fig:07}).
\item [$\bullet$] We used our data and those collected from the literature and the
web archives to investigate possible transit timing variations,
but we did not find any evidence of a departure from a linear
ephemeris.
\item [$\bullet$] We used the DFOSC, PEST and GROND data sets to revise the physical
parameters of the system. In particular, we obtained an estimation
of the radius of WASP-19\,b with a precision better than $2\%$,
which reveals that the planet is larger ($R_{\mathrm{b}}=1.410 \pm
0.017$) than reported in the discovery paper and in other studies
(see Table \ref{tab:09}). This implies that the density of
WASP-19\,b is lower than originally thought.
\item [$\bullet$] A joint analysis of our new photometric data and spectroscopic
results from the literature supports the picture that the WASP-19
system is roughly two times older than was originally estimated by
\citet{hebb2010}, in agreement with the more recent estimate of
\citet{tregloan2013}.
\item [$\bullet$] Thanks to the ability of GROND to measure stellar flux
simultaneously through seven different filters, covering quite a
large range of the optical and NIR windows, we measured a radius
variation of WASP-19\,b as a function of wavelength. We joined our
measurements with those obtained by other authors to reconstruct a
transmission spectrum of the planet's atmosphere in terms of
planet/star radius ratio. This transmission spectrum was compared
with two synthetic spectra, based on model atmosphere in chemical
equilibrium, where the PT profiles are assumed intermediate
between planet-wide and day-side, but with different opacity
characteristics (with and without gaseous TiO and VO). We found
that WASP-19\,b's atmosphere is presumingly dominated by
absorption by H$_{2}$O, Na and K, and no evidence for a strong
optical absorber at low pressure, which agrees well with the fact
that the atmosphere lacks a dayside inversion.
\item [$\bullet$] A joint analysis of the four occultations of WASP-19\,b was
performed and from this it was possible to measure an occultation
depth of $0.048 \pm 0.013 \%$, corresponding to an $i$-band
brightness temperature of $T_{\mathrm{B}} = 2544_{-110}^{+96}$\,K.
We combined this measurement with all others available in the
literature, providing a very wide coverage of the planet's
spectral energy distribution (750--8000\,nm). We compared these
experimental results with two dayside emission model spectra taken
from the literature and another two models constructed following
the methodology of \citet{fortney2008}, again considering both the
cases with and without TiO in the atmosphere of WASP-19\,b. The
model, corresponding to an atmosphere with a temperature of
$\Teff=2525$\,K, and no temperature inversion, returned the best
fit of the experimental measurements.
\item [$\bullet$] We conclude that WASP-19\,b seems to be inefficient at homogenizing
temperature, given how hot the dayside is. This agrees well with
the advective and radiative time-scale argument from
\citet{cowan2011}. For these very hot planets, the radiative time
is short, so advection fails to move energy fast enough to
homogenize the temperature.
\item [$\bullet$]
The planet probably does not fit the \citet{fortney2008} or
\citet{madhusudhan2012} classification schemes. It seems to be an
oxygen-rich planet (C/O $<$ 1), but with no TiO. A possible
explanation is that it is cold-trapped at depth.
\end{itemize}

\section*{Acknowledgements}
This paper is based on observations collected with the MPG/ESO
2.2\,m and the Danish 1.54\,m telescopes, both located at ESO La
Silla, Chile. Operation of the 2.2\,m telescope is jointly
performed by the Max Planck Gesellschaft and the European Southern
Observatory. Operation of the Danish telescope is based on a grant
to UGJ by the Danish Natural Science Research Council (FNU). GROND
was built by the high-energy group of MPE in collaboration with
the LSW Tautenburg and ESO, and is operated as a PI-instrument at
the 2.\,2m telescope. We thank Timo Anguita and R\'egis Lachaume
for technical assistance during the GROND observations.
The reduced light curves presented in this work will be made
available at the CDS ({\tt http://vizier.u-strasbg.fr/}) and at
{\tt http://www.astro.keele.ac.uk/$\sim$jkt/}.
J\,Southworth acknowledges financial support from STFC in the form
of an Advanced Fellowship.
The research leading to these results has received funding from
the European Community's Seventh Framework Programme
(FP7/2007-2013/) under grant agreement Nos.\ 229517 and 268421.
Funding for the Centre for Star and Planet Formation (StarPlan)
and the Stellar Astrophysics Centre (SAC) is provided by The
Danish National Research Foundation.
MD, MH, CL and CS acknowledge the Qatar Foundation for support
from QNRF grant NPRP-09-476-1-078.
SG and XF acknowledge the support from NSFC under the grant
No.\,10873031.
SG acknowledge the support from Chinese Academy of Sciences (grant
No. KJCX2-YW-T24).
The research is supported by the ASTERISK project (ASTERoseismic
Investigations with SONG and Kepler) funded by the European
Research Council (grant agreement No.\,267864).
OW (FNRS research fellow), FF (ARC PhD student), DR (FRIA PhD
student) and J\,Surdej acknowledge support from the Communaut\'e
fran\c{c}aise de Belgique - Actions de recherche concert\'ees -
Acad\'emie Wallonie-Europe.
TCH acknowledges financial support from the Korea Research Council
for Fundamental Science and Technology (KRCF) through the Young
Research Scientist Fellowship Program and is supported by the KASI
(Korea Astronomy and Space Science Institute) grant
2012-1-410-02/2013-9-400-00.
The following internet-based resources were used in research for
this paper: the ESO Digitized Sky Survey; the NASA Astrophysics
Data System; the SIMBAD data base operated at CDS, Strasbourg,
France; and the ar$\chi$iv scientific paper preprint service
operated by Cornell University.

%
%



\begin{thebibliography}{99}
%
\bibitem[Abe et al.(2013)]{abe2013}
Abe L. et al., 2013, A\&A, 553, A9
%
\bibitem[Anderson et al.(2010)]{anderson2010}
Anderson D.~R. et al., 2010, A\&A, 513, L3
%
\bibitem[Anderson et al.(2013)]{anderson2011}
Anderson D.~R. et al., 2013, MNRAS, 430, 3422
%
\bibitem[Ballerini et al.(2012)]{ballerini2012}
Ballerini P., Micela G., Lanza A.~F., Pagano I., 2012, A\&A, 539,
A140
%
\bibitem[Bean et al.(2013)]{bean2013}
Bean J.~L., D\'{e}sert J.-M., Seifahrt A., Madhusudhan N.,
Chilingarian I., Homeier D., Szentgyorgyi A., 2013, ApJ, 771, 108
%
\bibitem[Berdyugina(2005)]{berdyugina2005} %
Berdyugina S.~V., 2005, Living Rev. Solar Phys., 2, 8
%
\bibitem[Burton et al.(2012)]{burton2012}
Burton J.~R., Watson C.~A., Littlefair S.~P., Dhillon V.~S.,
Gibson N.~P., Marsh T.~R., Pollacco D., 2012, ApJS, 201, 36
%
\bibitem[Claret(2004)]{claret2004}
Claret A., 2004, A\&A, 428, 1001
%
\bibitem[Collier Cameron(1992)]{cameron1992} %
Collier Cameron A. 1992, Surface inhomogeneities on late-type
stars, ed. P.~B. Byrne, \& D.~J. Mullan (Springer, Berlin), 33
%
\bibitem[Copperwheat et al.(2013)]{copperwheat2013}
Copperwheat C.~M., et al., 2013, MNRAS, 434, 661
%
\bibitem[Cowan \& Agol(2011)]{cowan2011} %
Cowan N.~B., \& Agol E., 2011, ApJ, 729, 54
%
\bibitem[de Mooij et al.(2012)]{demooij2012}
de Mooij E.~J.~W. et al., 2012, A\&A, 538, A46
%
\bibitem[Dominik et al.(2010)]{dominik2010}
Dominik M. et al., 2010, AN, 331, 671
%
\bibitem[Doyle et al.(2013)]{doyle2013}
Doyle A.~P. et al., 2013, MNRAS, 428, 3164
%
\bibitem[Dragomir et al.(2011)]{dragomir2011}
Dragomir D. et al., 2011, ApJ, 142, 115
%
\bibitem[Enoch et al.(2010)]{enoch2010}
Enoch B., Collier Cameron A., Parley N.~R., Hebb L., 2010, A\&A,
516, A33
%
\bibitem[Fortney et al.(2005)]{fortney2005}
Fortney J.~J., Marley M.~S., Lodders K., Saumon D., Freedman R.,
2005, ApJ, 627, L69
%
\bibitem[Fortney et al.(2008)]{fortney2008}
Fortney J.~J., Lodders K., Marley M.~S., Freedman R.~S., 2008,
ApJ, 678, 1419
%
\bibitem[Fortney et al.(2010)]{fortney2010}
Fortney J. J., Shabram M., Showman A. P., Lian Y., Freedman R.~S.,
Marley M.~S., Lewis N.~K., 2010, ApJ, 709, 1396
%
\bibitem[Freedman et al.(2008)]{fr08}
Freedman R.~S., Marley M.~S., Lodders K., 2008, ApJ, 174, 513
%
\bibitem[Fukui et al.(2013)]{fukui2013}
Fukui A. et al., 2013, ApJ, 770, 95
%
\bibitem[Gibson et al.(2008)]{gibson2008}
Gibson N. P., Pollacco D., Simpson E.~K., et al., 2008, A\&A, 492,
603
%
\bibitem[Gibson et al.(2010)]{gibson2010}
Gibson N.~P. et al., 2010, MNRAS, 404, L114
%
\bibitem[Gillon et al.(2006)]{gillon2006}
Gillon M., Pont F., Moutou C., Bouchy F., Courbin F., Sohy S.,
Magain P., 2006, A\&A, 459, 249
%
\bibitem[Greiner et al.(2008)]{greiner2008}
Greiner J. et al., 2008, PASP, 120, 405
%
\bibitem[Harps{\o}e et al.(2013)]{harpsoe2013}
Harps{\o}e K.~B.~W. et al., 2013, A\&A, 549, A10
%
\bibitem[Hebb et al.(2010)]{hebb2010}
Hebb L. et al., 2010, ApJ, 708, 224
%
\bibitem[Hellier et al.(2011)]{hellier2011}
Hellier C., Anderson D.~R., Collier Cameron A., Miller G.~R.~M.,
Queloz D., Smalley B., Southworth J., Triaud A.~H.~M.~J., 2011,
ApJL, 730, L31
%
\bibitem[Lendl et al.(2012)]{lendl2012}
Lendl M., Gillon M., Queloz D., Alonso R., Fumel A., Jehin E.,
Naef D., 2013, A\&A, 552, A2
%
\bibitem[Lodders \& Fegley(2002)]{Lodders02}
Lodders K., Fegley B., 2002, Icarus, 155, 393
%
\bibitem[Mancini et al.(2013a)]{mancini2013a}
Mancini L. et al., 2013a, A\&A, 551, A11
%
\bibitem[Mancini et al.(2013b)]{mancini2013b}
Mancini L. et al., 2013b, MNRAS, 430, 2932
%
\bibitem[Mohler-Fischer et al.(2013)]{mohler2013}
Mohler-Fischer M. et al., 2013, A\&A, 558, A55
%
\bibitem[Nikolov et al.(2012)]{nikolov2012}
Nikolov N., Henning Th., Koppenhoefer J., Lendl M., Maciejewski
G., Greiner J., 2012, A\&A, 539, 159
%
\bibitem[Nikolov et al.(2013)]{nikolov2013}
Nikolov N., Chen G., Fortney J.~J, Mancini L., Southworth J., van
Boekel R., Henning Th., 2013, A\&A, 553, A26
%
\bibitem[Madhusudhan(2012)]{madhusudhan2012}
Madhusudhan N., 2012, ApJ, 758, 36
%
\bibitem[Mandel \& Agol(2002)]{mandel2002} %
Mandel K., Agol E., 2002, ApJL, 580, L171
%
\bibitem[Pierini et al. (2012)]{pierini2012}
Pierini D. et al., 2012, A\&A 540, A45
%
\bibitem[Press et al.(1992)]{press1992}%
Press W.~H., Teukolsky S.~A., Vetterling W.~T., Flannery B.~P.,
1982, Numerical Recipes in FORTRAN (Cambridge: Cambridge Univ.
Press)
%
\bibitem[Rabus et al.(2009)]{rabus2009} %
Rabus M. et al., 2009, A\&A, 494, 391
%
\bibitem[Sanchis-Ojeda \& Winn(2011)]{sanchis2011}
Sanchis-Ojeda R., Winn J.~N., 2011, ApJ, 743, 61
%
\bibitem[Schwarz (1978)]{Schwarz1978}%
Schwarz G.~E., 1978, Annals of Statistics, 6, 461
%
\bibitem[Silva(2003)]{silva2003} %
Silva A.~V.~R., 2003, ApJL, 585, L147
%
\bibitem[Sing et al.(2011)]{sing2011} %
Sing D.~K. et al., 2011, MNRAS, 416, 1443
%
\bibitem[Southworth(2008)]{southworth08} %
Southworth J., 2008, MNRAS, 386, 1644
%
\bibitem[Southworth(2010)]{southworth10}
Southworth J., 2010, MNRAS, 408, 1689
%
\bibitem[Southworth(2011)]{southworth11}
Southworth J., 2011, MNRAS, 417, 2166
%
\bibitem[Southworth(2012)]{southworth12}
Southworth J., 2012, MNRAS, 426, 1291
%
\bibitem[Southworth et al.(2004)]{southworth2004}
Southworth J., Maxted P.~F.~L., Smalley B., 2004, MNRAS, 349, 547 %
%
\bibitem[Southworth et al.(2009a)]{southworth2009a}
Southworth J. et al., 2009, MNRAS, 396, 1023
%
\bibitem[Southworth et al.(2009b)]{southworth2009b}
Southworth J. et al., 2009b, MNRAS, 399, 287
%
\bibitem[Southworth et al.(2012a)]{southworth2012a}
Southworth J., Bruni I., Mancini L., Gregorio J., 2012a, MNRAS,
420, 2580
%
\bibitem[Southworth et al.(2012b)]{southworth2012b}
Southworth J., Mancini L., Maxted P.~F.~L., Bruni I.,
Tregloan-Reed J., Barbieri M., Ruocco N., Wheatley P.~J., 2012b,
MNRAS, 422, 3099
%
\bibitem[Southworth et al.(2012c)]{southworth12c}
Southworth J., et al., 2012, MNRAS, 426, 1338
%
\bibitem[Stetson(1987)]{stetson1987}
Stetson P.~B., 1987, PASP, 99, 191
%
\bibitem[Strassmeier(2009)]{strassmeier2009}
Strassmeier K.~G., 2009, Astron. Astrophys. Rev., 17, 251
%
\bibitem[Tregloan-Reed et al.(2013)]{tregloan2013}
Tregloan-Reed J., Southworth J., Tappert C., 2013, MNRAS, 428,
3671
%
\bibitem[Vogt et al.(1999)]{vogt1999} %
Vogt S.~S., Hatzes A., Misch A., K\"{u}rster M., 1999, ApJS, 121,
547
%
\bibitem[Winn et al.(2008)]{winn2008}%
Winn J.~N. et al., 2008, ApJ, 683, 1076
%
%
\end{thebibliography}
\end{document}